\documentclass[12pt,preprint]{aastex}
\begin{document}

\newcommand{\dif}{\mathrm{d}}
\newcommand{\mclust}{\texttt{mclust}}
\newcommand{\optim}{\texttt{optim}}
\newcommand{\sparr}{\texttt{sparr}}
\newcommand{\spatstat}   {\texttt{spatstat}}
\newcommand{\mlpowerlaw}   {\texttt{ml\_powerlaw}}

\newcommand{\todo}[1]       {\noindent[{\bf \footnotesize #1}]} \newcommand{\tbr}[1]        {#1}
\newcommand{\tbd}           {\_\_\_\_  {\bf (TBD)}}     \newcommand{\revise}[1]     {{#1}}

\newcommand{\mystix} {MYStIX}
\newcommand{\Spitzer} {{\em Spitzer}}
\newcommand{\Herschel} {{\em Herschel}}
\newcommand{\Chandra} {{\em Chandra}}
\newcommand{\ACIS}    {{ACIS}}
\newcommand{\CIAO}    {{\em CIAO}}
\newcommand{\Sherpa}  {{\em Sherpa}}
\newcommand{\Chart}   {{\em ChaRT}}
\newcommand{\SAOTrace}{{\em SAOTrace}}
\newcommand{\DSnine}  {{\em DS9}}
\newcommand{\MARX}    {{\em MARX}}
\newcommand{\AEacro}  {{\em AE}}
\newcommand{\TARA}    {{\em TARA}}
\newcommand{\XSPEC}   {{\em XSPEC}}
\newcommand{\FTOOLS}  {{\em FTOOLS}}
\newcommand{\HEASOFT} {{\em HEASOFT}}
\newcommand{\IDL}     {{\em IDL}}
\newcommand{\Rlan}   {{\em R}}

\shorttitle{Intrinsic Populations of Young Stellar Clusters}
\shortauthors{Kuhn et al.}
\slugcomment{Accepted for publication in ApJ, January 12, 2015}

\title{The Spatial Structure of Young Stellar Clusters. II. Total Young Stellar Populations}

\author{Michael A. Kuhn\altaffilmark{1,2,3}, Konstantin V. Getman\altaffilmark{1}, Eric D. Feigelson\altaffilmark{1}} 
\altaffiltext{1}{Department of Astronomy \& Astrophysics, 525 Davey Laboratory, Pennsylvania State University, University Park, PA 16802, USA}
\altaffiltext{2}{Instituto de Fisica y Astronom\'{i}a, Universidad de Valpara\'{i}so, Gran Breta\~{n}a 1111, Playa Ancha, Valpara\'{i}so, Chile}
\altaffiltext{2}{Millennium Institute of Astrophysics}

\begin{abstract}
We investigate the intrinsic stellar populations (estimated total numbers of OB and pre--main-sequence stars down to 0.1~$M_\odot$) that are present in 17 massive star-forming regions (MSFRs) surveyed by the MYStIX project. The study is based on the catalog of $>$31,000 MYStIX Probable Complex Members with both disk-bearing and disk-free populations, compensating for extinction, nebulosity, and crowding effects. Correction for observational sensitivities is made using the X-ray Luminosity Function (XLF) and the near-infrared Initial Mass Function (IMF)---a correction that is often not made by infrared surveys of young stars. The resulting maps of the projected structure of the young stellar populations, in units of intrinsic stellar surface density, allow direct comparison between different regions. Several regions have multiple dense clumps, similar in size and density to the Orion Nebula Cluster. The highest projected density of $\sim$34,000~stars~pc$^{-2}$ is found in the core of the RCW~38 cluster. Histograms of surface density show different ranges of values in different regions, supporting the conclusion of Bressert et al. (2010, B10) that no universal surface-density threshold can distinguish between clustered and distributed star-formation. However, a large component of the young stellar population of MSFRs resides in dense environments of 200--10,000~stars~pc$^{-2}$ (including within the nearby Orion molecular clouds), and we find that there is no evidence for the B10 conclusion that such dense regions form an extreme ``tail'' of the distribution.
Tables of intrinsic populations for these regions are used in our companion study of young cluster properties and evolution.
\end{abstract}

\section{Introduction}

The Milky Way Galaxy is a critical part in the universe for studying star formation. Only here can the populations of low-mass stars---making up the vast majority of stars---be resolved and the full spatial structure of young stellar clustering and molecular clouds be analyzed, revealing detailed information about how star formation progresses within a region. Most stars, including the Sun \citep{Gounelle12,Dukes12}, are born in clusters with OB-type stars, so it is important to study the massive star-forming regions (MSFRs) in the solar neighborhood. The young stellar clusters in these regions can be precursors to open clusters, but most of their stars become gravitationally unbound due to gas expulsion, so an understanding the star-formation histories and early cluster dynamics in these regions provides clues about how bound clusters and field stars are produced \citep[e.g.,][]{Goodwin06,Pfalzner11,Kruijssen12b,Banerjee14}.

Historically, studies of massive Galactic star-forming regions have been hindered by difficulties inherent to Galactic Plane astronomy; in particular, field stars greatly outnumber members of the star-forming region in optical or infrared (IR) images \citep[e.g.,][]{King13,Kuhn13b}. X-ray surveys readily detect star-forming region members due to the high X-ray luminosities from strong magnetic activity of pre-main-sequence stars ($L_X\sim10^{-3}L_{\mathrm{bol}}$; Preibisch et al.\ 2005a), and these surveys are not strongly effected by nebulosity, obscuration, or crowding. 
Excess IR emission from disk-bearing young stars has proven to be another useful method of establishing membership, but IR-only surveys will miss the large populations of members without dusty protoplanetary disks and such studies rarely account for observational sensitivities in determining intrinsic stellar populations, as we attempt to do in this paper. The combination of X-ray selected stars and IR-excess selected stars can provide better samples of stars in MSFRs than either method alone \citep{Feigelson13,Townsley11}. Thus, the analysis of empirical distributions of inferred stellar mass and X-ray luminosities from the combined samples can be used to estimate total populations \citep[e.g.,][and references therein]{Getman12}.  

The {\bf M}assive {\bf Y}oung Star-Forming Complex {\bf St}udy in {\bf I}nfrared and {\bf X}-ray \citep[\mystix;][]{Feigelson13} examines 20 nearby massive star-forming regions (MSFRs) using a combination of archival \Chandra\ X-ray imaging, 2MASS+UKIDSS near-IR (NIR), and \Spitzer\ mid-IR (MIR) survey data. The catalogs of young stars include both high-mass and low-mass stars, and disk-bearing and disk-free stars \citep{Broos13}. We use this sample of stars to study the intrinsic populations of young stars across 17 of the MYStIX MSFRs.

The present study is closely based on the constructions of the MYStIX Probable Complex Members catalog \citep[MPCM;][]{Broos13} and the statistical segregation of MPCMs into $\sim$140 subclusters by \citet[][Paper~I]{Kuhn14}. Our principal objective here is to overcome sensitivity limitations of the MPCM catalog for each MYStIX MSFR in order to normalize the observed stellar distributions to intrinsic stellar distributions.  We obtain two quantities of interest:  the total intrinsic stellar population and the stellar surface densities in each MYStIX subcluster.  The total populations are important inputs into a multivariate analysis of young cluster properties in our forthcoming study (Kuhn et al.\ 2014b, Paper~III).  The stellar surface densities address long-standing issues about typical environments in which stars form.

\subsection{Thresholds for Clustered Star Formation}

The statistical relationships that traditionally underlay our understandings of star-formation processes have been scale-free relationships like the \citet{Salpeter55} stellar initial mass function (IMF) and the Kennicutt--Schmidt law relating global galactic star formation rate to interstellar material \citep{Schmidt59,Kennicutt12}. Nevertheless, important preferred scales for star formation were later found. The power-law IMF for high stellar masses peaks around 0.2-0.3 $M_\odot$ and declines for lower mass stars (Chabrier 2003). And, in the Galactic neighborhood, a threshold for star formation was found at $A_V \approx 7$ magnitudes dust absorption ($\approx n[H_2] \sim 3\times10^4$ cm$^{-3}$) associated with Galactic disk stability 
\citep{Johnson04,Lada10,Martin01,Schaye04,Leroy08},
 although this rule is not applicable within the inner 0.5~kpc of the Galaxy where star-formation is comparatively suppressed \citep{Longmore13}.

The surface density of stellar populations in star-forming regions, in units of stars per square parsec, has been a property of interest for the field of star-cluster formation \citep[e.g.,][]{Carpenter00,Lada03,Allen07,Jorgensen08,Gutermuth09}. The surface densities of young stars can also have astrophysical implications, such as tidal truncation of protoplanetary disks \citep{Pfalzner05}, binary star distributions \citep{Bate09a,Moeckel11}, or the survival of clusters after molecular cloud dispersal \citep{Kruijssen12a}. 
\citet[][henceforth B10]{Bressert10} have recently examined the shape of the distribution of stellar surface densities in star-forming regions. They use samples of disk/envelope-bearing stars identified through IR excess in regions within 0.5~kpc of the Sun, including the Gould Belt \citep{Allen06}, the Orion A and B molecular clouds \citep{Megeath12}, the Taurus molecular cloud \citep{Rebull10}, and the regions from the cores-to-disks \citep[c2d;][]{Evans03} project---thus their sample is dominated by low-mass young-stellar-object (YSO) environments. B10 argue that if ``clustered'' star-formation and ``distributed'' star-formation were two distinct star-formation processes, then ``clustered'' and ``distributed'' populations should appear as distinct modes in the surface-density distribution \citep[cf.][]{Gieles12,Pfalzner12}, and that such a scenario could be tested by searching for a ``scale'' of star formation separating these two modes. Such a kink in the surface-density distribution has been used in some investigations of the structure of young stellar clusters, for example by \citet{Gutermuth09}, to separate distinct clusters of stars in star-forming regions. Given that B10 find a smooth distributions of surface density from their data, which they report is adequately fit by a log-normal distribution with a peak at 22~stars~pc$^{-2}$, they conclude that no such ``scale'' exists, at least for low-mass YSO environments present in the solar neighborhood. 

Nevertheless, the log-normal distribution from B10 is not scale free, but instead a peak at 22~stars~pc$^{-2}$ and width of 0.85~dex suggests a density distribution weighted towards low-density environments. If this result is a general characteristic of most star formation in the Galaxy, rather than just the nearby regions investigated by B10, it would have implications for theories of star formation as numerous researchers have discussed. For example, \citet{Parmentier13} find that their models of local-density-driven star formation from a single molecular clump could produce the 22~stars~pc$^{-2}$ scale from B10. \citet{King12} suggest that surface densities significantly greater than 22~stars~pc$^{-2}$ could indicate that a cluster has undergone a ``cool collapse phase.'' \citet{Kruijssen12b} present a model in which the low fraction of star-formation that results in bound clusters is, in part, a result of a density spectrum weighted towards low surface densities. \citet{Parker11} notes that dynamical processing of primordial binaries by clusters depends on whether most stars form in low density regions as suggested by B10, or whether higher density clusters are more common. And, \citet{deJO} investigate the threshold densities in star-forming regions where stellar interactions affect habitable planet formation---and the fraction of the stars born in environments above their $3\times10^3$~stars~pc$^{-2}$ threshold would depend on whether high-density regions are just a tail of the B10 log-normal or a different mode not seen in the B10 sample. B10 support this interpretaton of the empirical results from their sample stating, ``only a small fraction ($<$26 per cent) of stars form in dense clusters where their formation and/or evolution is expected to be influenced by their surroundings.''

It is important to investigate whether these results continue to hold for more massive star-forming regions (those regions containing O-type stars). Nearly 70\% of B10's sample comes from the Orion Giant Molecular Clouds, which do contain O-type stars, but they note that the IR-excess methods used are ineffective at identifying young stars in the presence of nebulosity and crowding in this complex. Their other star-forming regions are lower mass. Given that studies of the mass function for star-forming regions favor the birth of stars in more massive complexes, investigation of these complexes in a way that is more effective at probing the densest regions could be helpful for determining the validity of B10's suggestion that clustered stars ($\Sigma>200$~stars~pc$^{-2}$) exist in the tail of the surface-density distribution, rather than being a dominant component. Nevertheless, a definitive study of surface-density distributions for star formation would require the construction of an unbiased survey of all star-forming environments. \mystix\ neither includes the most massive star-forming environments in the Galaxy (such as W~43, Wd~1, NGC~3603, or the Arches Cluster) nor includes large angular area studies of diffuse molecular clouds necessary to capture the lowest surface density environments. 

\subsection{MYStIX}

The MYStIX survey \citep{Feigelson13} differs from many previous studies in that it focuses on relatively massive star-forming regions lying in nearby Galactic spiral arms, and supplements samples of IR-excess young stars with X-ray selected young stars and spectroscopically identified OB stars.  

\mystix\ is a survey of 20 of the nearest ($d<3.6$~kpc) MSFRs that have been observed with NASA's {\it Chandra X-ray Observatory}, the {\it Spitzer Space Telescope}, and the United Kingdom Infra-Red Telescope (UKIRT) or the Two Micron All Sky Survey \citep[2MASS;][]{Skrutskie06}. The MYStIX regions include the Orion Nebula, the Flame Nebula, W~40, RCW~36, NGC~2264, the Rosette Nebula, the Lagoon Nebula, NGC~2362, DR~21, RCW~38, NGC~6334, NGC~6357, the Eagle Nebula, M~17, W~3, W~4, the Carina Nebula, the Trifid Nebula, NGC~3576, and NGC~1893, from which a sample of 31,784 MYStIX Probable Complex Members \citep[MPCMs;][]{Broos13} is obtained. The MPCM catalog thus consists of young stars that are X-ray selected, IR excess selected, or OB stars from the literature. MYStIX provides the cleanest and largest lists of young stars for most of the 20 regions included in the study, so these catalogs should be scientifically useful for different purposes. One of the requirements of the MYStIX project was to use sensitive and homogeneous data analysis procedures for all 20 regions to facilitate inter-comparisons between regions. Special procedures had to be developed to deal with challenges working in the Galactic Plane, as described in the MYStIX technical-catalog papers: \citet{Kuhn13a}, \citet{Townsley14}, \citet{King13}, \citet{Kuhn13b}, \citet{Naylor13}, \citet{Povich13}, and \citet{Broos13}.

The spatial distributions of MPCMs in 17 of the \mystix\ MSFRs are investigated in \citet[][henceforth Paper~I]{Kuhn14} and in this work. Three regions are omitted, NGC~3576, W~3, and W~4, because they lack $JHK$ UKIRT photometry and have a low match rate between X-ray sources and sources from the 2MASS catalog. We use a subset of the MPCM sources ($\sim$17,000 stars) produced after X-ray selected MPCMs are pruned to a uniform X-ray sensitivity within each region (Paper~I). This eliminates artificial surface density gradients associated with differing X-ray exposure times in \Chandra\ mosaics and the sensitivity variations within each \Chandra\ field due to telescope coma and vignetting, the ``egg-crate effect'' \citep{Townsley11}. To prune a region, we remove sources with X-ray photon fluxes \citep[{\it log\_PhotonFlux\_t};][]{Broos13} that are lower than the completeness limits provided in Table~1 of Paper~I.

Nevertheless the resulting observed surface densities, used by the analysis in Paper~I, do not contain the entire intrinsic population, differ in sensitivity from MSFR to MSFR, and are affected by spatially variable $N_H$ absorption and mid-IR sensitivity.  As pre--main-sequence (PMS) X-ray luminosities strongly scale with stellar mass \citep{Telleschi07}, inconsistent X-ray sensitivities due to differing \Chandra\ exposure times and MSFR distances lead to different samplings of the cluster IMFs.  We overcome these selection effects calibrating the observed X-ray luminosity function (XLF) and IMF distributions to the Orion Nebula Cluster (ONC) that serves as a template for young cluster populations, rather than attempting to model instrumental and observational effects.

The organization of this paper (Paper~II) is as follows. We analyze the IMF and X-ray luminosity function (XLF) to infer intrinsic populations from observed young stellar populations (Section~2). We derive intrinsic stellar surface density maps from these populations (Section~3), and investigate surface density distributions (Sections~4 and~5). The \mystix\ sample of star-forming regions are typically richer than those in the sample studied by B10. The \mystix\ MSFRs exhibit a large diversity in their surface density distributions (ranging from $\sim$10 to $\sim$30,000~stars~pc$^{-2}$), neither showing a tendency to follow a universal surface density distribution nor showing a convincing peak at some characteristic surface density. These results are discussed in Sections~4 and 5 and summarized in  the conclusion (Section~6).

\section{Stellar Populations}

The completeness limits and detection fractions\footnote{The completeness limit of a sample is defined to be the minimum luminosity (or mass) such that nearly 100\% of objects with greater luminosity (or mass) are included in the sample. The detection fraction is the number of observed sources $N_\mathrm{obs}$ divided by the intrinsic number of sources $N_\mathrm{tot}$.}
of the MPCM samples vary from region to region, due to differences in distance, obscuration, and X-ray and IR observation exposures. Extrapolations of the total numbers of stars in a region, which we infer empirically from observed MPCM samples based on comparisons with the ONC, are necessary for comparison of intrinsic properties of stellar populations in different star-forming regions.

\subsection{X-ray Luminosity Functions}

Young stars in the \Chandra\ Orion Ultradeep Project \citep[COUP;][]{Getman05} had large numbers of X-ray counts, allowing X-ray luminosities ($L_\mathrm{t,c}$; {\bf t}otal 0.5--8.0~keV band and absorption {\bf c}orrected) to be obtained by parametric modeling of the X-ray spectrum using the XSPEC code \citep{Arnaud96}. As the \mystix\ stars are mostly too faint for this procedure, X-ray luminosities for other MPCMs were computed using non-parametric calibrations for PMS stars \citep[XPHOT;][]{Getman10} by \citet{Kuhn13a}, \citet{Townsley14}, \citet{Broos11}, and \citet{Kuhn10}. 

The probability distribution of $L_\mathrm{t,c}$, called the XLF, is associated with the IMF due to the statistical link between X-ray luminosity and stellar mass $M$ \citep[e.g.,][]{Feigelson93,Preibisch05,Telleschi07}. The assumption of a ``universal XLF'' \citep{FeigelsonGetman05} has been used to estimate total populations in young stellar clusters, including Cep~B \citep{Getman06}, M~17 \citep{Broos07}, NGC~6357 \citep{Wang07}, Rosette \citep{Wang08,Wang09,Wang10}, W~40 \citep{Kuhn10}, Trumpler~15 \citep{Wang11}, Trumpler~16 \citep{Wolk11}, Sh~2-254/255/256/257/258 \citep{Mucciarelli11}, NGC~1893 \citep{Caramazza12}, and IC~1396 \citep{Getman12}. During PMS stellar evolution, there is a weak relation between X-ray luminosity and age \citep[e.g.,][]{PreibischFeigelson05, Pandey14}; however, $L_\mathrm{t,c}$ does not rapidly decline during the first 5~Myr, unlike the rapid decrease in bolometric luminosity $L_\mathrm{bol}$ during PMS evolution along the Hayashi track. Instead, $L_\mathrm{t,c}\sim M$ appears to be the fundamental relationship rather than $L_\mathrm{t,c}\sim L_\mathrm{bol}$ \citep{Getman14b}. Thus, X-ray luminosity evolution appears to have little effect on the shape of the PMS XLF \citep[e.g.,][]{Bhatt13}. 

Following these previous studies, we use a sample of stars from COUP to approximate the probability distribution of the universal XLF. 
The COUP study contains a sample of 839 lightly absorbed stars \citep{Feigelson05}, which are identified as the members of the ONC, while the more highly absorbed stars are identified as being embedded in the Orion Molecular Cloud (OMC) behind the ONC. These lightly obscured COUP stars are complete down to a mass of 0.1--0.2~$M_\odot$ \citep[with partial coverage into the proto--brown-dwarf regime;][]{Preibisch05b} and show an XLF shape characterized by a falling distribution at high luminosities with a break to an approximately flat distribution at luminosities below $L_\mathrm{t,c} \approx 10^{30.4}$~erg~s$^{-1}$. Henceforth, we label this distribution the COUP XLF. The tail with X-ray luminosities greater than this turnover can be fit with a power-law (Pareto) distribution of slope $\alpha$, with a minimum variance unbiased estimator $\alpha^*$ and variance $\it{Var}(\alpha^*)$ given by the equations
\begin{equation}
\alpha^* = -\left(1-\frac{2}{n}\right)\frac{n}{\sum_{i=1}^n\left(\ln x_i - \ln x_m\right)}~\mathrm{and}~\it{Var}(\alpha^*) = \frac{\alpha^{*2}}{n-3},
\end{equation}
where $x_m$ is the X-ray luminosity of the turnover point and $x_i$ is the X-ray luminosity of the $i$th source in a sample of $n$ sources in the distribution tail \citep{Johnson94}. Thus, $\alpha^* = -0.9\pm0.1$ with $n=61$, while the distribution is roughly flat in logarithmic bins below the turnover point. \citet{Mucciarelli11} have also found similar $L_\mathrm{t,c}$ Pareto distributions for the ONC and the Sh~2-254--258 regions.

The $L_\mathrm{t,c}$ distributions for the \mystix\ regions can be compared to the empirical distribution function (EDF) of 839 lightly absorbed COUP sources to further investigate the universality of the XLF shape. For example, \citet[][their Figure~9b]{Wang08} shows that the XLF cumulative distributions in the Rosette star-forming region agree with the COUP XLF above the X-ray luminosity completeness limit for the Rosette sample. For this analysis it is necessary to be cautious about how completeness limits are treated because differential absorption can change the apparent shape of the XLF if the sample of highly absorbed sources is incomplete. For example, a sample of sources that are more deeply embedded will have a higher mean luminosity than a sample of sources from the same observation that are unabsorbed, which could lead to a flattening of the power-law of the combined distribution. 

\begin{figure}
\centering
\includegraphics[angle=0.,width=3.0in]{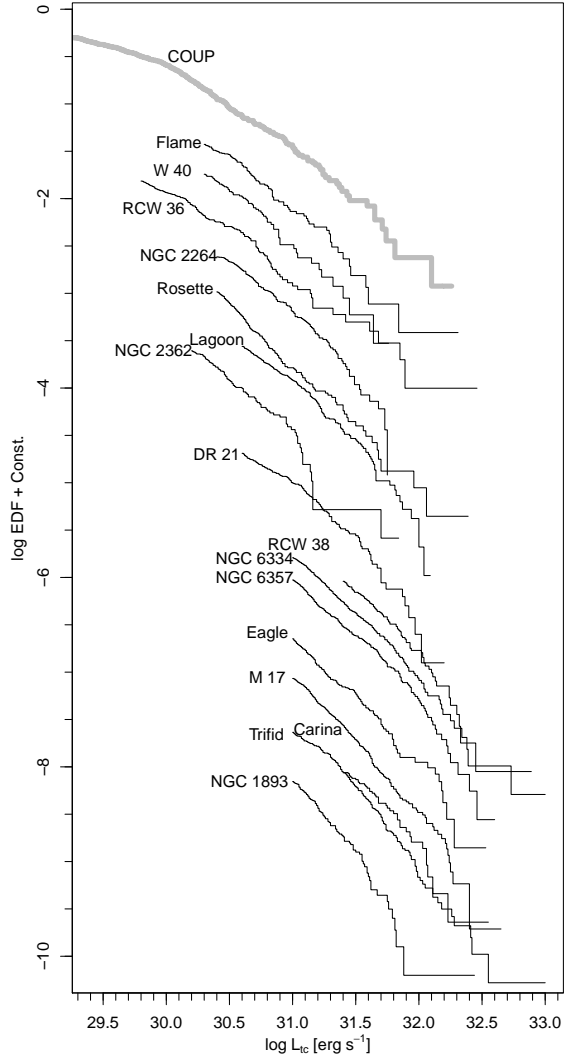}
\caption{The EDF of X-ray luminosities, $L_{tc}$, for the 839 lightly obscured, low-mass stars from COUP (complete down to 0.1-0.2~$M_\odot$) is shown by the thick gray line. The black lines, from top to bottom, are the X-ray luminosity EDFs for the other \mystix\ regions. Vertical shifts of 0.3 units are used to separate the different lines for visual clarity. Lines end at the completeness limit for the full sample (i.e.\ the completeness limit for the most heavily obscured subpopulation in the region).
\label{full_xlfs.fig}}
\end{figure}

In Figure~\ref{full_xlfs.fig}, we show the COUP EDF running from unity at very low luminosities ($L_\mathrm{t,c}\sim 10^{27}$~erg~s$^{-1}$) to zero for the most luminous PMS star in the Orion Nebula field. Note that spectroscopically identified OB stars have been removed from both the COUP XLF and from \mystix\ samples considered here because \mystix\ regions can differ widely in their massive stellar subpopulations. The $L_\mathrm{t,c}$ EDFs for the other regions are shown below, with arbitrary vertical spacings. Only the portion of the XLF where the sample is complete is shown. (The completeness limit for the full sample is set to the completeness limit for the most heavily obscured subpopulation in the region.) In general there is excellent agreement between the shapes of the different lines. Some curvature can be seen in the COUP XLF between $10^{30.5}$~erg~s$^{-1}$ and $10^{32.5}$~erg~s$^{-1}$, which is also reflected in the shapes of the XLFs for other regions. The nearest regions tend to probe a lower-luminosity section of the XLF, while the more distant regions tend to probe a higher-luminosity section of the XLF. Due to the XLF curvature, the XLF shape appears less steep for nearer regions and steeper for more distant regions. Table~\ref{full_xlfs.tab} (Column~2) gives the power-law indices\footnote{The power-law fits are often poor and we do not recommend that these values be used for astrophysical interpretation.} (for the full sample) calculated over the regions shown in Figure~\ref{full_xlfs.fig}. This confirms the trend in which more distant regions have steeper slopes---not because of differences in intrinsic XLF shape, but due to differences in the available portion of the XLF.

\begin{deluxetable}{lrrrrr}
\tablecaption{XLF properties of MSFRs \label{full_xlfs.tab}}
\tablewidth{0pt}
\tablehead{
\colhead{Region} & \colhead{$\alpha^*$} & \colhead{$N_\mathrm{tot}$} & \colhead{$N_{ME<1.5}$} & \colhead{$N_{1.5<ME<2.5}$} &  \colhead{$N_{ME>2.5}$} \\
\colhead{} & \colhead{} & \colhead{(stars)} & \colhead{(stars)} & \colhead{(stars)} & \colhead{(stars)}\\
\colhead{(1)} & \colhead{(2)} & \colhead{(3)} & \colhead{(4)} & \colhead{(5)} & \colhead{(6)}
}
\startdata
Orion 	& -0.9 & 2600 & 700 & 1200 & 720\\
Flame 	& -1.0 & 800 & 55 & 250 & 500\\
W 40 	& -1.1 & 520 & 0 & 250 & 270\\
RCW 36    & -0.8 &550 & 0 & 230 & 320\\
NGC 2264 & -1.0 & 1900 & 1100 & 230 & 570\\
Rosette	& -1.3 & 2500 & 840 & 1000 & 610\\
Lagoon	& -1.1 & 3800 & 2000 & 870 & 930\\
NGC 2362	& -1.1 & 600 & 570 & 30 & 0\\
DR 21	& -1.0 & 2900 & 50 & 550 & 2253\\
RCW 38	& -1.3 & 9900 & 70 & 1900 & 8000\\
NGC 6334	& -1.2 & 9400 & 180 & 2900 & 6200\\
NGC 6357	& -1.2 & 12000 & 440 & 7400 & 3700\\
Eagle	& -1.3 & 8100 & 1300 & 4300 & 2500\\
M 17		& -1.3 & 16000 & 170 & 5400 & 11000\\
Carina	& -1.5 & 34000 & 9500 & 17000 & 7000\\
Trifid     	& -1.0 & 3100 & 1500 & 1100 & 470\\
NGC 1893 & -1.5 & 4600 & 1700 & 2400 & 590\\
\enddata
\tablecomments{
Column~1: Name of \mystix\ MSFR. Column~2: Power-law index ($\dif \log N / \dif \log L_{t,c}$) for the portion of the XLF tail shown in Figure~\ref{xlf.fig} (uncertainties are $\pm$0.1). Column~3: Inferred intrinsic total population. Columns~4--6: Inferred intrinsic numbers of stars with $ME<1.5$~keV, $1.5<ME<2.5$~keV, and $ME>2.5$~keV, respectively. (Due to rounding, the sum of Columns~4, 5, and 6 is not always equal to Column~3.)
}
\end{deluxetable}

\subsubsection{Intrinsic Numbers of Stars \label{strata.sec}}

If we accept the assumption of a ``universal XLF'' with the COUP sample serving as a template, the total stellar population of a star-forming region may be extrapolated from the missing stars at low luminosities where the region's observed XLF has dropped to zero due to incompleteness, but where the existence of these stars may be inferred from the universal XLF shape. To perform this calculation, the COUP XLF histogram is scaled so that it matches the observed XLF in the section of the XLF where the observed sample is complete. The completeness limit for the full sample of young stars will be the completeness limit for the most absorbed subpopulation of stars in the region. However, few of the most deeply embedded protostars may be observed, so this population will be poorly characterized, and the inferred total populations may be considered lower limits that do not necessarily account for all undetected, embedded stars. 

If we further assume that X-ray absorptions from the molecular cloud are independent of the intrinsic stellar X-ray luminosities, we can estimate the intrinsic numbers of stars with different amounts of absorption in the star-forming region---this is useful because it allows us to compare total populations in more extinguished parts of a star forming region to total populations in less extinguished areas. Using this assumption, any subset of stars selected by X-ray absorption will be drawn independently from the ``universal XLF'' and therefore have the same XLF shapes---allowing us to perform the XLF population analysis described above on the subset. This assumption may not be entirely true; for example, mass segregation may cause more X-ray luminous stars to lie preferentially in the center of a cluster where absorptions are often higher. There is also weak evidence for a factor of $\sim$2 systematic increase in X-ray luminosity from the younger Class~I to the older Class~II--III systems \citep{Prisinzano08}. Nevertheless, the absorption stratified MYStIX XLFs all show consistency with the COUP XLF, indicating that these are not major effects.

Interstellar medium absorption of MPCMs may be evaluated using $J-H$ color indices or X-ray Median Energy \citep[$ME$;][]{Getman10} indicators, both of which increase as absorbing columns increase \citep{Getman14b}. The spread in absorptions for subclusters of stars is provided in Paper~I (their Figure 8), which shows the median $J-H$ and $ME$ for clusters of young stars in the \mystix\ regions. There is a clump of data points with $ME \approx 1.5$~keV (the unembedded population), while the absorbed population ranges over $1.5<ME<5.0$~keV. The bulk of the absorbed groups of stars are moderately absorbed, with $1.5<ME<2.5$~keV, and tend to have physical properties similar to the unabsorbed clusters. In contrast, the most absorbed groups of stars are much more compact ($r_\mathrm{core}\approx0.08$~pc) and tend to be centered on dense molecular filaments or clumps \citep{Getman14b}. 

\begin{figure}
\centering
\includegraphics[angle=0.,width=6.0in]{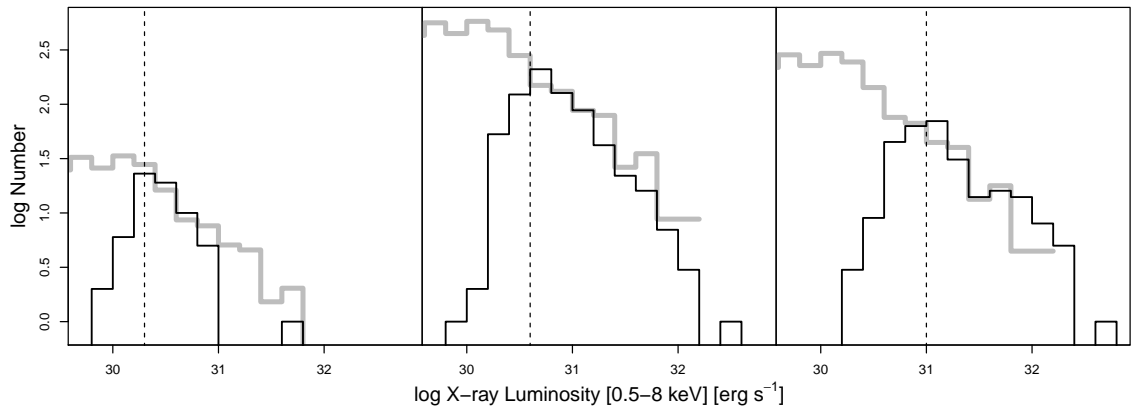}
\caption{The XLF for NGC~6357 X-ray selected MPCMs with (left) $ME<1.5$~keV, (center) $1.5<ME<2.5$~keV, and (right) $ME>2.5$~keV. Stars with known spectral types earlier than B3 are not included. The COUP XLF of 839 lightly absorbed stars is scaled to these populations. Completeness limits for the three absorption strata are indicate by the vertical dashed line. \label{xlf.fig}
(The complete figure set (17 images) is available in the online journal)
}
\end{figure}

We choose to stratify the \mystix\ MPCM stars into three absorption strata using $ME$ divisions at 1.5 and 2.5~keV. This captures different aspects of the spatial structure of the stars: roughly, the population of stars outside the molecular cloud, the population within the molecular cloud, and the population associated with dense molecular filaments and cores.  
Figure~\ref{xlf.fig} shows the observed XLF histograms for these three subpopulations in NGC~6357 (black lines): the $ME<1.5$~keV sources (left), the $1.5<ME<2.5$~keV sources (center), and the $ME>2.5$~kev sources (right). The completeness limits in this region vary from $10^{30.3}$, to $10^{30.6}$, to $10^{31.0}$, for these three strata, respectively. The template COUP XLF (gray lines) is scaled to these three populations, using a vertical shift that minimizes the area between the two EDFs in the range where the XLF is approximately complete.  
The XLFs of all three strata show agreement between the shape of the sample XLF and the COUP XLF. 
The intrinsic number of stars in each absorption stratum can be estimated by integrating the number of stars under the scaled universal XLF curve, i.e.\ the gray COUP line. Thus, the inferred intrinsic population in NGC~6357 is $\sim$440 lightly obscured stars, $\sim$7400 moderately obscured stars, and $\sim$3700 highly obscured stars.\footnote{We repeated this analysis using alternate $ME$ divisions to investigate whether these choices have any systematic affect on the total populations of stars that we infer. Using  1.0, 1.25, 1.5, 1.75, and 2.0~keV for the first division and 2.0, 2.25, 2.5, 2.75, and 3.0~keV for the second division, we find that inferred total numbers of stars (Table~1, column 3) only change by $\sim$3\%, and there is no systematic trend towards over or underestimation.} Similar plots of stratified XLFs are provided for the other 16 regions as a figure set in the electronic version of the article.

As expected, for every region, the more highly absorbed strata have a higher $L_\mathrm{t,c}$ completeness limit than the less absorbed strata. Generally, there is good agreement with shape of the COUP XLF; however, in most cases the sample becomes incomplete before reaching the turnover point at $L_\mathrm{t,c} \approx 10^{30.4}$~erg~s$^{-1}$ in the XLF. The Flame Nebula is one of the few cases where the completeness limit is less luminous than the XLF turnover point in the lightly and moderately absorbed strata, so the ``flat'' portions of the XLFs can be compared---the figure indicates that the XLFs are consistent. There is also indication of this turnover in the lightly absorbed stratum for NGC~2264.  

The extrapolated intrinsic stellar populations for each stratum in each region are provided in Table~\ref{full_xlfs.tab}. These values are combined to produce estimates of the total intrinsic populations for each star-forming region (Column~3). These values of $N_\mathrm{tot}$ are the principal empirical results of this study. Effects of distance and observational sensitivity should be approximately corrected by this analysis, but inferred total populations for more distant, less complete regions will be less precise and may be affected by additional systematics. To investigate these effects, we simulate the X-ray sensitivity for the Orion Nebula region if it were at the distance of the Carina Nebula ($\sim$2.3~kpc) with the pruned X-ray sensitivity ($log\_PhotonFlux\_t > -5.9$) for the \Chandra\ Carina Complex Project \citep[CCCP;][]{Townsley11}. \citet[][their Table~3]{Getman05a} report 21 close doubles with $<$2$^{\prime\prime}$ separations in COUP, which would be indistinguishable at the distance of Carina. For this test, we combine their X-ray luminosities into a single source for the purpose of X-ray sensitivity limits and XLF analysis. From the 120 X-ray sources in this reduced-sensitivity Orion sample (out of 1216 original X-ray sources), we infer a total of $\sim$2000 stars rather than $\sim$2600 stars. Thus our $N_\mathrm{tot}$ values may underestimate the true populations by up to $\sim$30\%, with most of the missing stars from the highly obscured $ME$ stratum. Effects of distance and observational sensitivity are mentioned in Sections~3 and~4.

\clearpage\clearpage

\setcounter{figure}{1}
\begin{figure}
\centering
\includegraphics[angle=0.,width=2.4in]{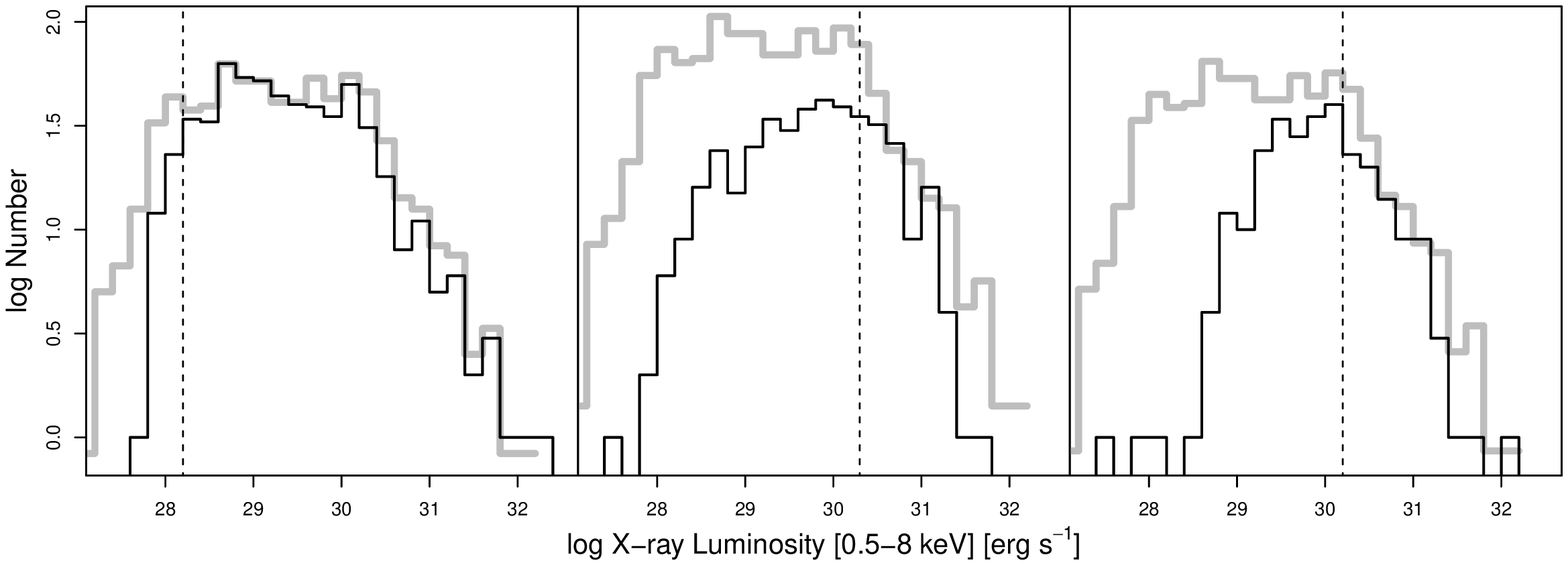}
\includegraphics[angle=0.,width=2.4in]{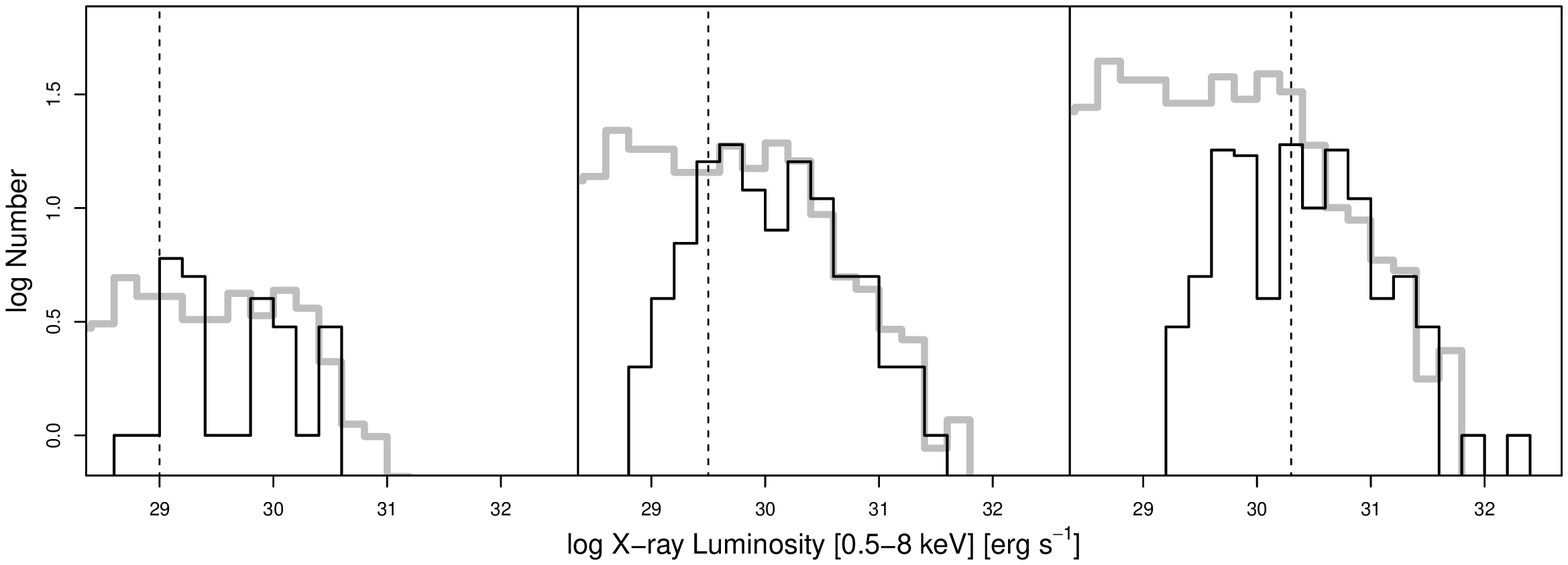}\\
\includegraphics[angle=0.,width=2.4in]{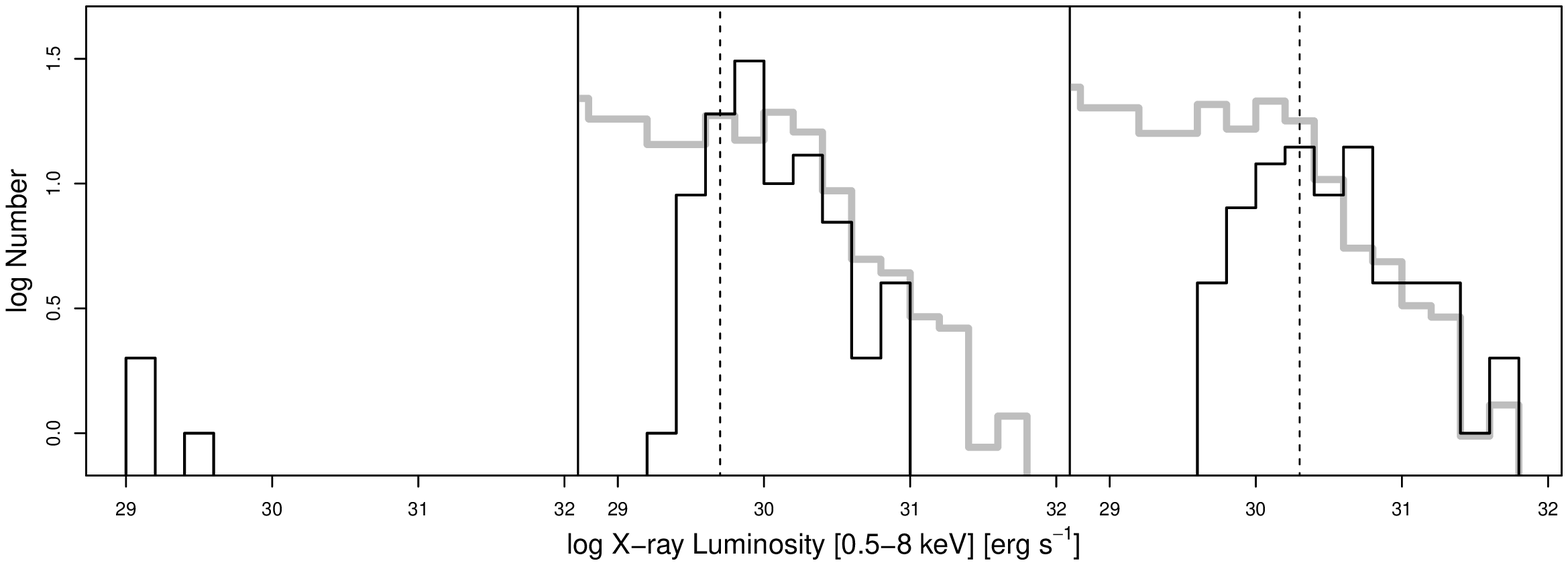}
\includegraphics[angle=0.,width=2.4in]{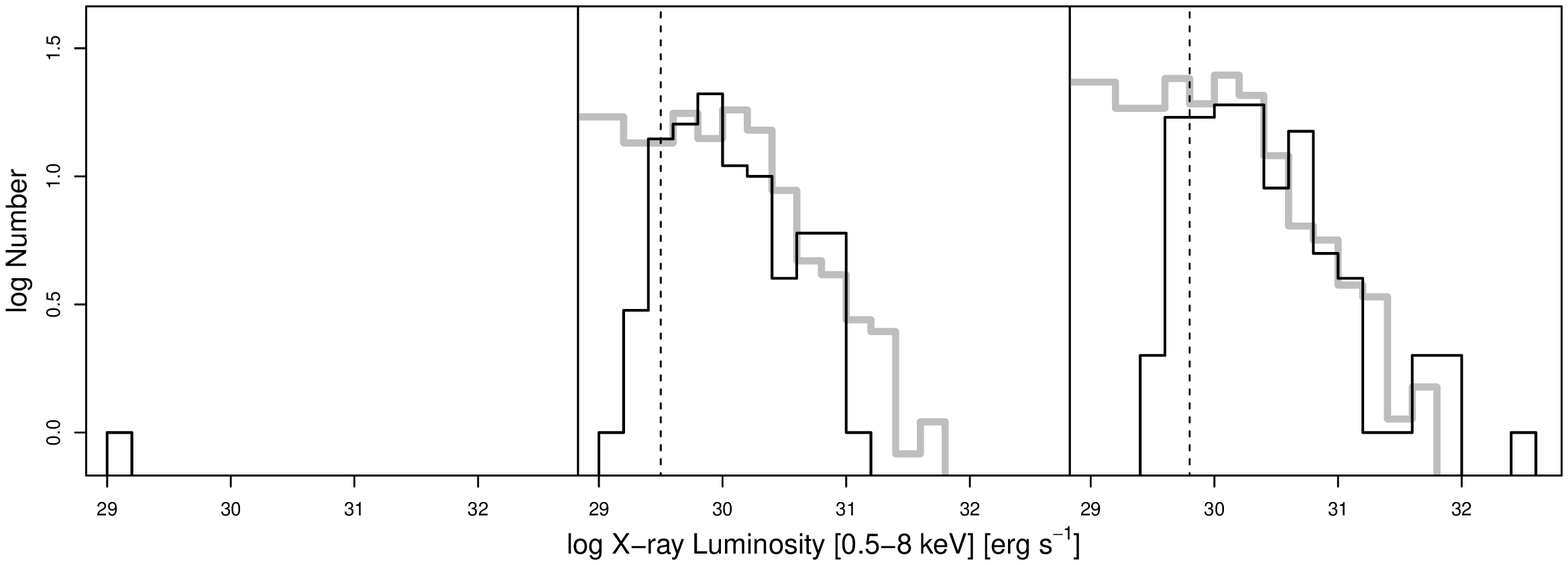}\\
\includegraphics[angle=0.,width=2.4in]{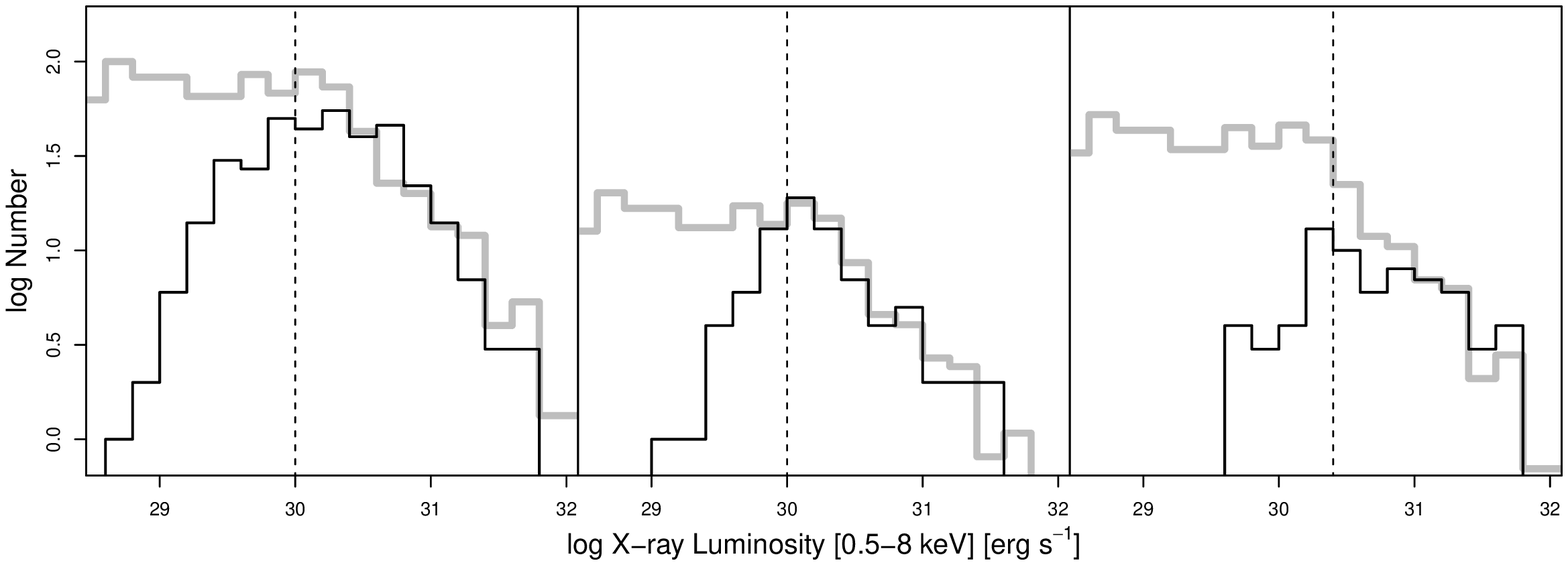}
\includegraphics[angle=0.,width=2.4in]{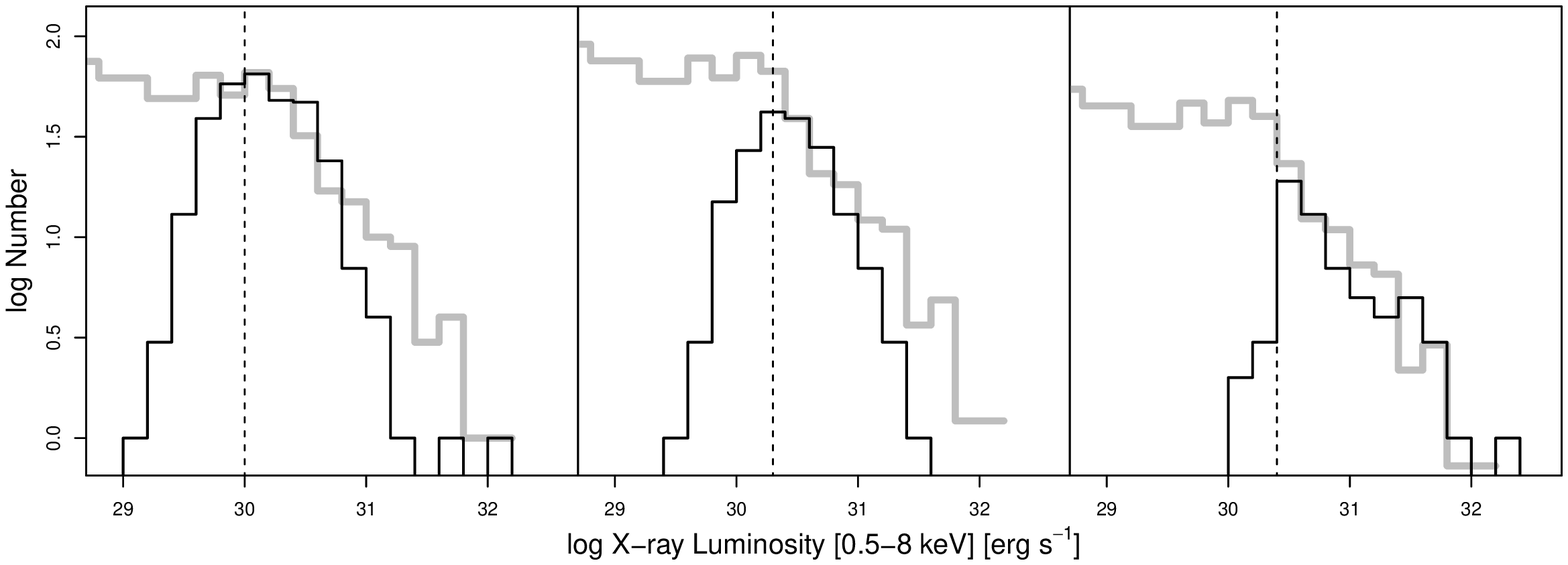}\\
\includegraphics[angle=0.,width=2.4in]{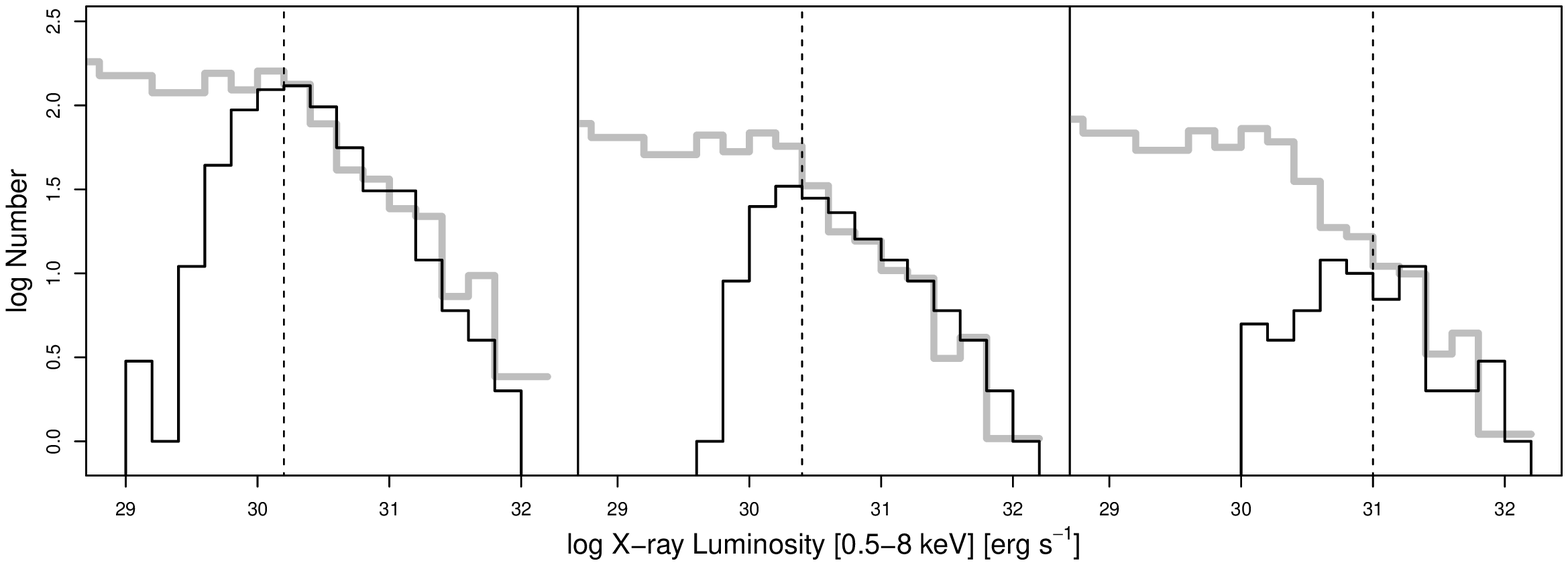}
\includegraphics[angle=0.,width=2.4in]{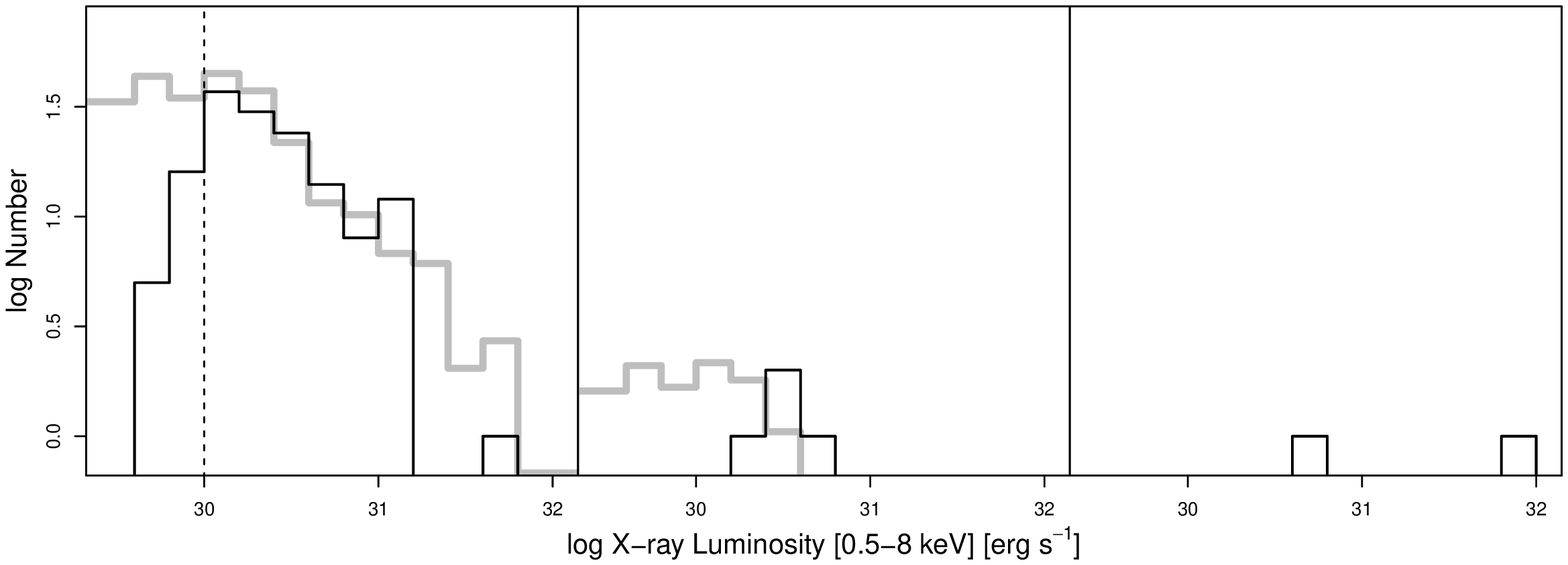}\\
\includegraphics[angle=0.,width=2.4in]{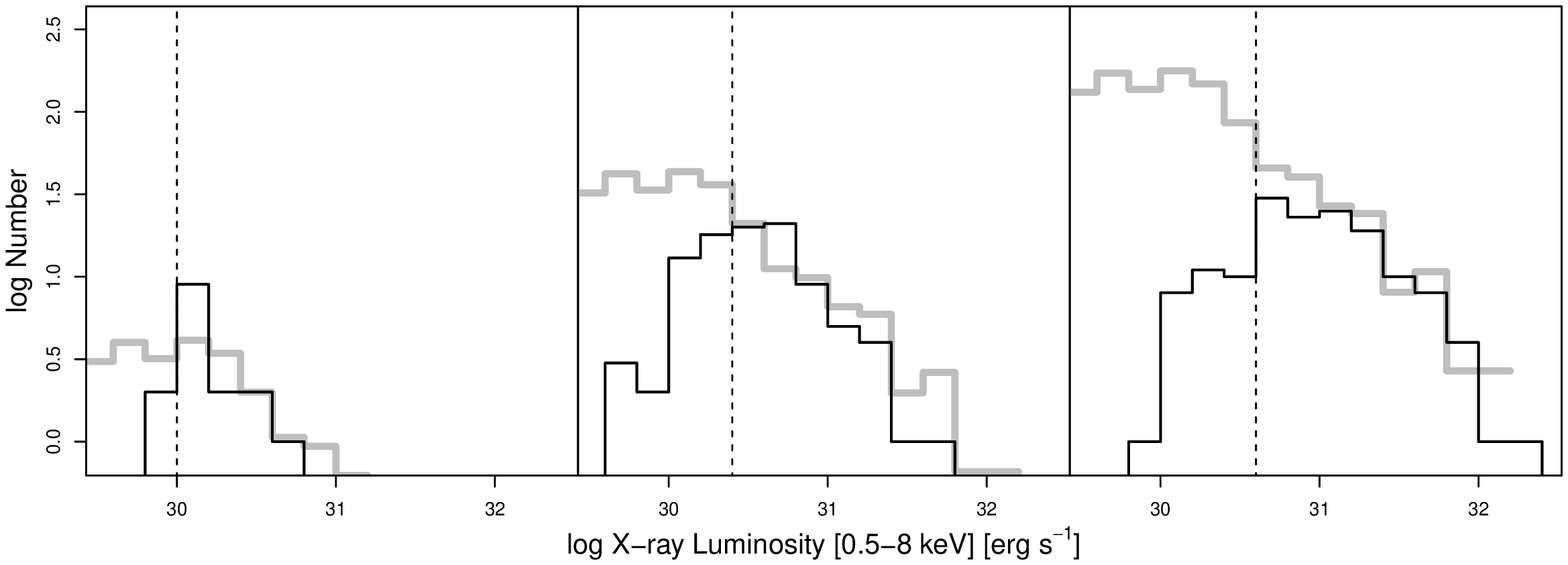}
\includegraphics[angle=0.,width=2.4in]{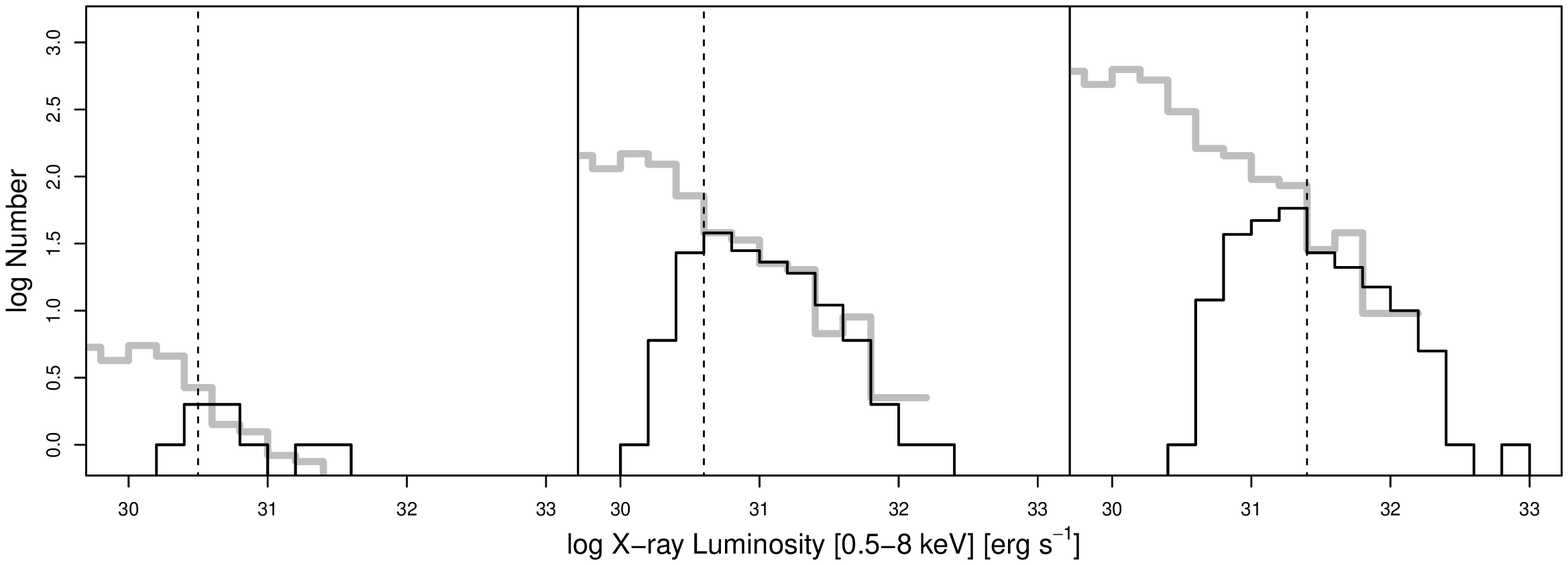}\\
\includegraphics[angle=0.,width=2.4in]{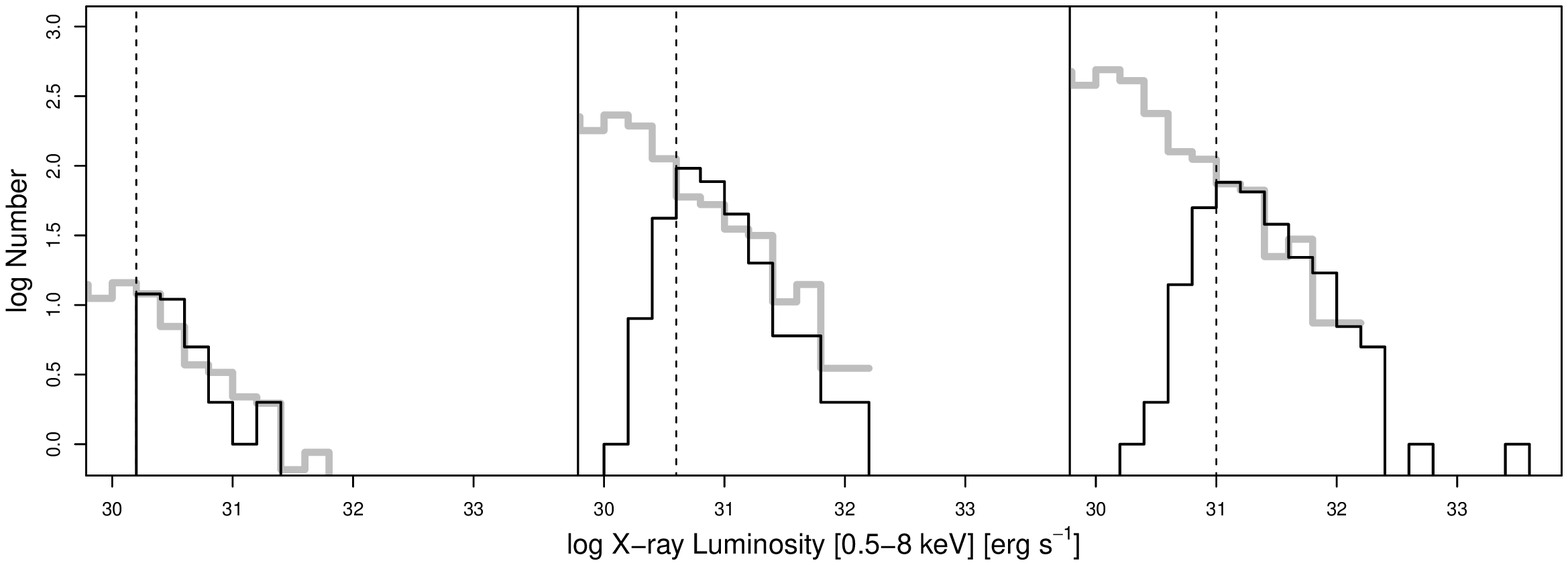}
\includegraphics[angle=0.,width=2.4in]{./f2l.eps}\\
\includegraphics[angle=0.,width=2.4in]{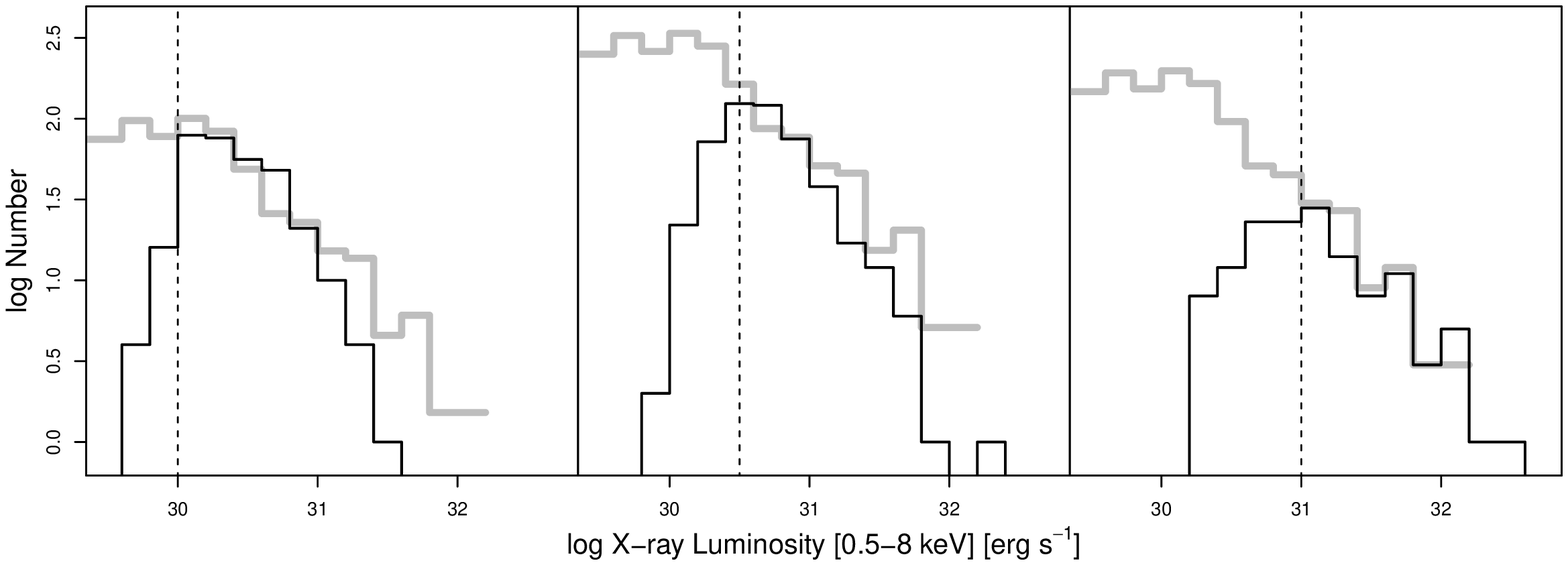}
\includegraphics[angle=0.,width=2.4in]{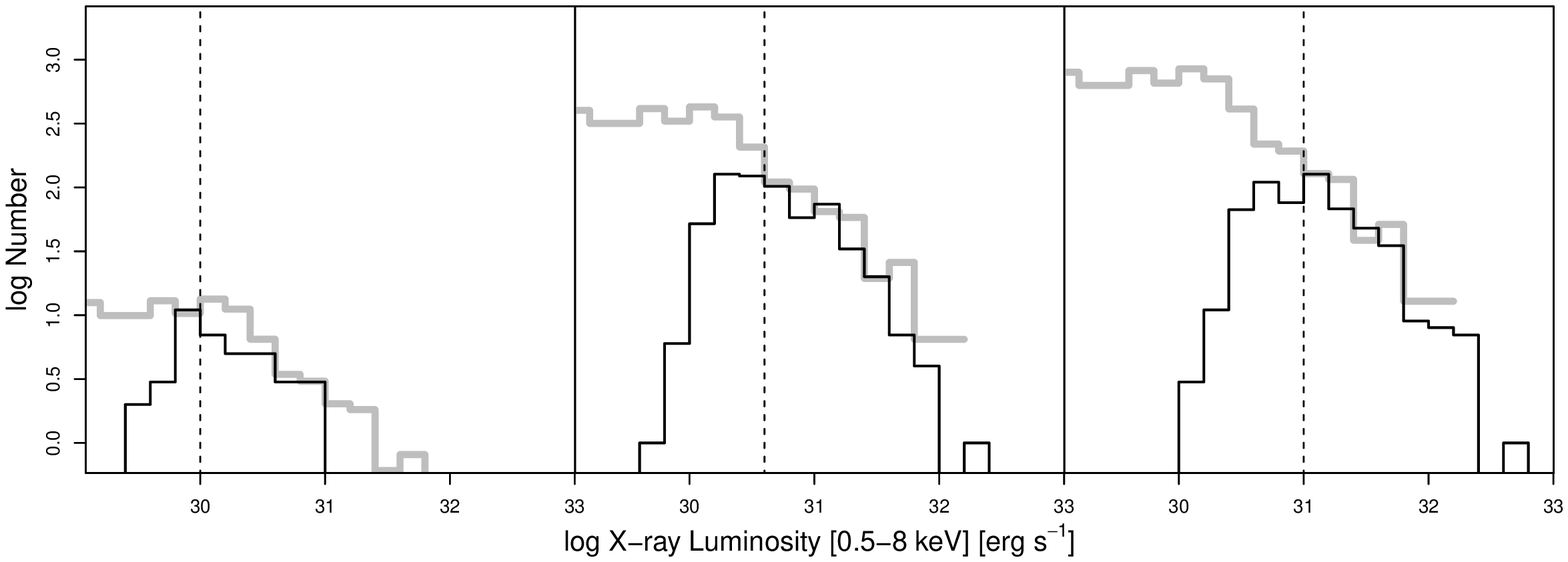}\\
\includegraphics[angle=0.,width=2.4in]{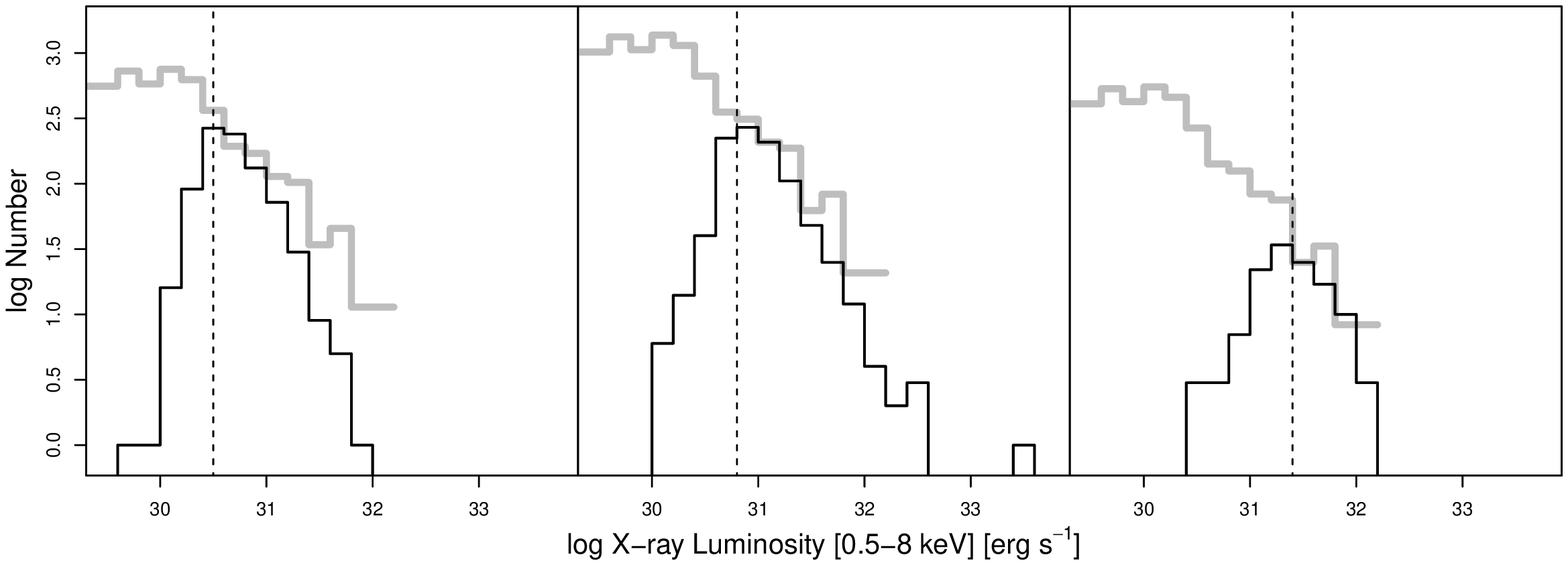}
\includegraphics[angle=0.,width=2.4in]{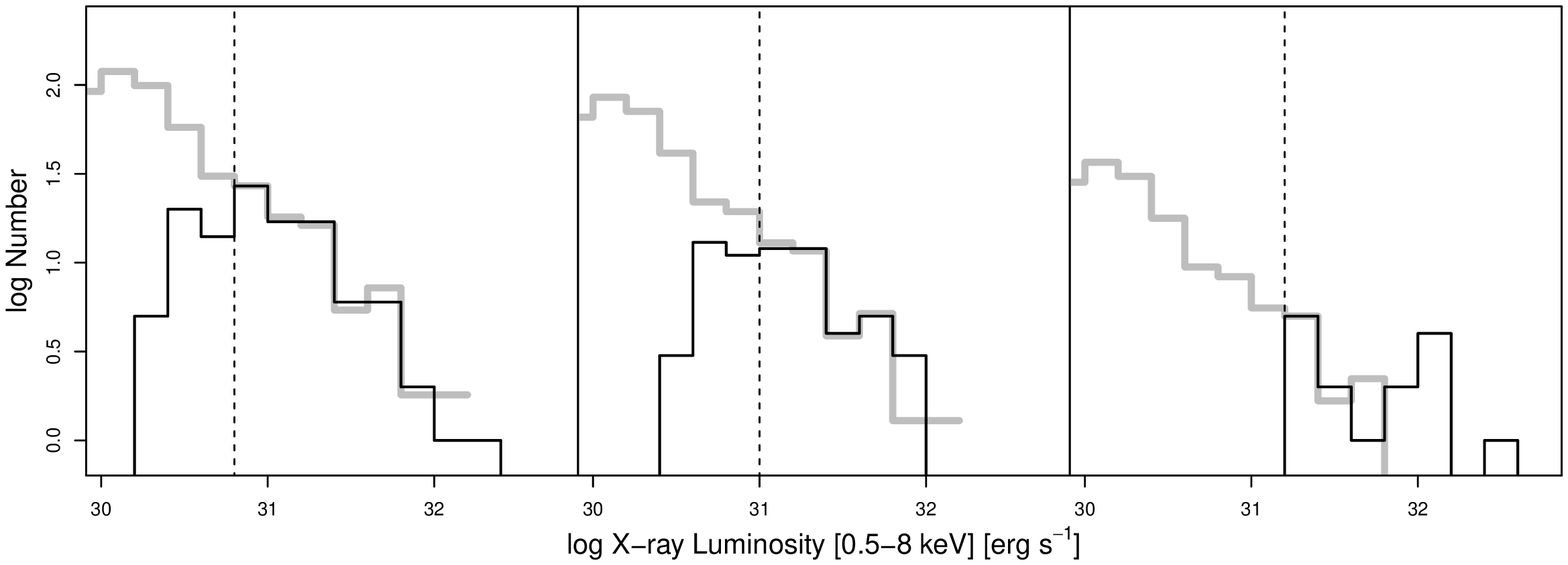}\\
\includegraphics[angle=0.,width=2.4in]{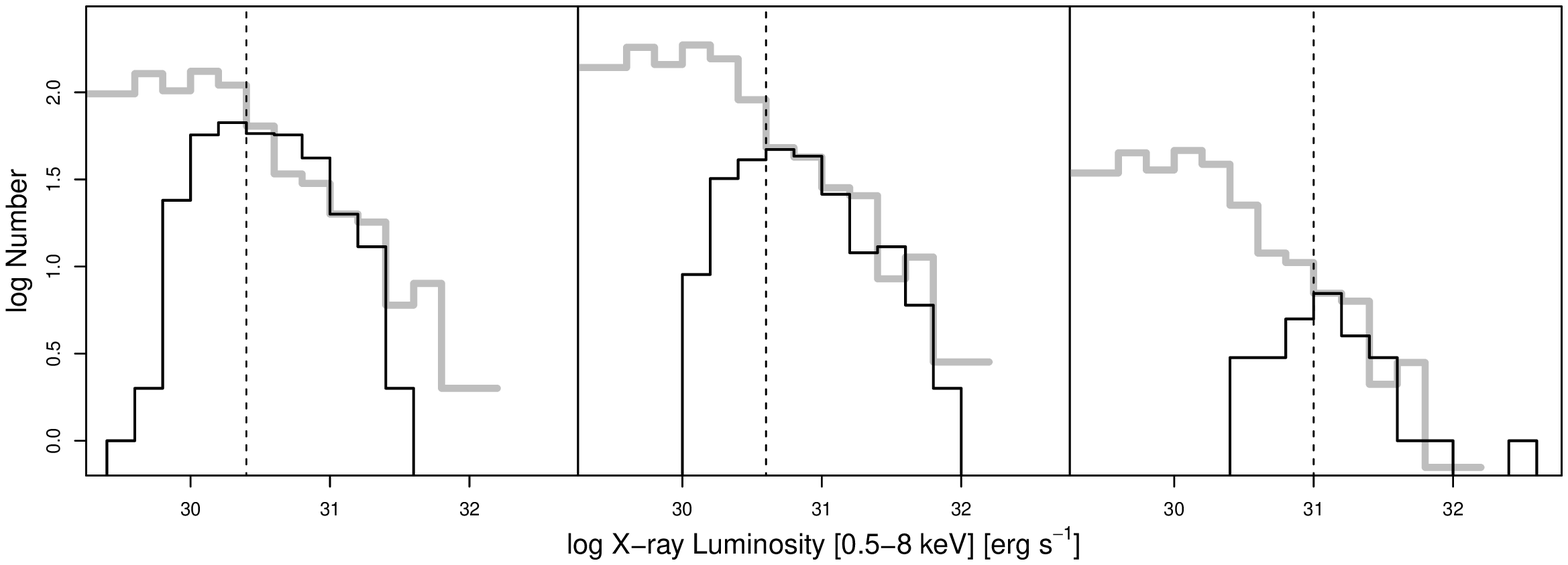}
\caption{(Left to right and top to bottom) Orion, Flame, W~40, RCW~36, NGC~2264, Rosette, Lagoon, NGC~2362, DR~21, RCW~38, NGC~6334, NGC~6357, Eagle, M~17, Carina, Trifid, NGC~1893}
\end{figure}
\clearpage\clearpage

\subsubsection{Detection Fraction for X-ray Selected MPCMs \label{detection.sec}}

The MPCM X-ray detection fraction, $f = N_\mathrm{obs}/N_\mathrm{tot}$, can be used as a correction factor to convert observed surface densities of young stars into intrinsic surface densities. However, dividing the surface density of a region by a single detection fraction does not account for spatial variation in sensitivity. The statistical sample of X-ray selected young stars provided in Paper~I, is a pruned subset of the MPCMs to which a uniform photon-flux limit is applied to correct for observational sensitivity effects, including \Chandra\ telescope vignetting, variation in point-spread function, and differing net exposure times in a mosaicked field. However, their samples do not control for differing luminosity completeness limits due to variable extinction. 

An improved estimate of stellar surface densities can be made using the absorption-stratified samples from Section~\ref{strata.sec}. 
Each $ME$ stratum has a narrower range of absorptions than the full sample, so the spatial variation in sensitivity within these samples will be lower than for the full sample. The sources in each stratum are adaptively smoothed (described below) to produce surface density maps, and the detection fraction for each stratum given by the XLF analysis.

The spatially dependent detection fractions for the full Paper~I X-ray selected samples are computed using the equation,
\begin{equation}
f_\mathrm{full}(\alpha,\delta) = \frac{f_1\Sigma_1(\alpha,\delta) + f_2\Sigma_2(\alpha,\delta)+f_3\Sigma_3(\alpha,\delta)}{\Sigma_1(\alpha,\delta) + \Sigma_2(\alpha,\delta)+\Sigma_3(\alpha,\delta)},
\end{equation}
where $f_i$ is the detection fraction of the $i$th stratum and $\Sigma_i(\alpha,\delta)$ is the adaptively smoothed observed surface density of that stratum. 
Detection fraction tends to be roughly constant across the fields of view, with dips found where molecular clouds and cores are located. But, detection fraction varies strongly between regions, so it is essential that scientific comparison between regions be based on the intrinsic young stellar surface densities rather than observed MPCM surface densities.\footnote{The alternative $ME$ strata (the two divisions ranging from 1.0--2.0 and 2.0--3.0~keV) produce maps with the same overall morphology as those reported here, without much change in completeness fraction over most of the field of view, but with 10--15\% difference at the extreme values of the map.}

\begin{figure}
\centering
\includegraphics[angle=0.,width=6.0in]{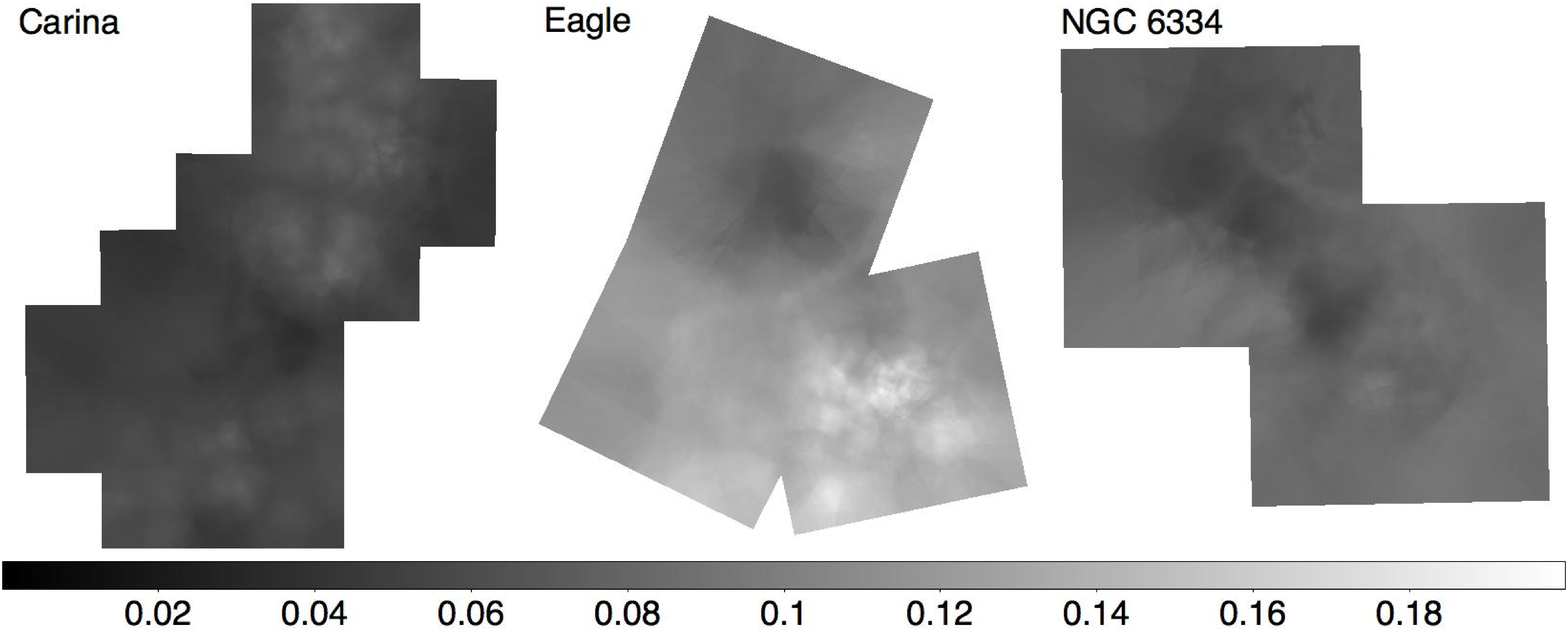}
\caption{The inferred fractional completeness of the MPCM catalogs in various regions. The fraction of young-stellar members that we expect to include in our MPCM list at any point is indicated by the shading of the maps, with dark shades indicating low detection fractions and light shades indicating high detection fractions as shown on the colorbar. (The data behind this figure is provided with the electronic edition of this article.)
\label{completenessfractionmap.fig}}
\end{figure}

Figure~\ref{completenessfractionmap.fig} shows detection fraction maps for the Carina, Eagle, and NGC 6334 fields of view. The larger distance and shorter \Chandra\ exposure times for the Carina Nebula compared to the Eagle Nebula result in a lower overall detection fraction for Carina compared to Eagle. In the Eagle Nebula, the bubble around the main NGC~6611 cluster results in a relatively high detection fraction, while the embedded subclusters to the north-east have lower detection fractions. A molecular filament passes through NGC~6334 from north-east to south-west, producing a notable trough in the detection fraction map of this region. Overall, detection fractions for MYStIX MSFRs range between 1--60\%.

Table~\ref{imf_xlf.tab} lists the intrinsic stellar populations inferred from the X-ray MPCM populations for the 142 subclusters in Paper~I, counting stars out to 4 times the subcluster core radius.\footnote{The numbers of observed stars reported in Paper~I count all the stars from the center of a subcluster out to 4 core radii. We use this cutoff because the projected half-mass radii of the subclusters are difficult to measure. If numbers for other radii are desired, they can be calculated from these data using Equation~3 in Paper~I to obtain a correction factor. We define $r_4=4.0\times r_c$, where $r_c$ is the ``isothermal ellipsoid'' core radius in Paper~I.} 
The detection fractions for these subclusters were obtained by interpolating the detection fraction at the location of each star assigned to a subcluster and taking the mean value, weighted to correct for incompleteness of the sample. The intrinsic population of a subcluster is the number of observed stars in that subcluster from Paper~I, multiplied by the fraction of those stars that are X-ray selected, and divided by the detection fraction of the X-ray selected sample. 

\clearpage\clearpage

\subsection{Initial Mass Functions}

The $J$~vs.~$J-H$ diagram has been used to obtain mass estimates of young stars, and thereby IMFs for young clusters, in a number of multiwavelength studies \citep[e.g.,][]{Getman08,Kuhn10,Wright10}. 
For low mass stars, individual inferred masses typically have 30\% systematic errors \citep[][their Appendix A]{Kuhn10}; however, this uncertainty is relatively small compared to the range of stellar masses in these samples and unimportant when the scientific questions concern their collective distributions. The method may be biased for the youngest stars: the spectral energy distribution modeling by \citet{Povich13} indicates that NIR absorption from a heavy disk or envelope may substantially increase the $J$-band magnitude, causing a star to appear to have a lower mass than it truly does. This method does not account for the mass of circumstellar material, which is substantial for protostars. And, furthermore, this method is insensitive to very young protostars undetected in NIR bands.

For each MYStIX subcluster, we adopt the median age from \citet{Getman14b}---their $Age_{JX}$ estimates where these exist, otherwise their $Age_{JH}$ estimates---and use the \citet{Siess97} numerical pre-main sequence evolutionary models to estimate absorption and mass for each star. A completeness mass-limit is estimated empirically from the inferred stellar masses for each subcluster. The \citet{Maschberger13} IMF is then scaled to the complete end of the observed mass functions, and the number of missing stars down to 0.1--0.2~$M_\odot$ is extrapolated. 

Table~\ref{imf_xlf.tab} lists the intrinsic numbers of stars calculated from IMF analysis for each of the 142 Paper~I subclusters. While there are numerous potential sources of error in this analysis, at a minimum there is a counting-statistics uncertainty of $N_\mathrm{tot}/\sqrt{N_{M>Mlim}}$, where $N_\mathrm{tot}$ is the inferred intrinsic population and $N_{M>Mlim}$ is the number of stars used to scale the IMF. 

\begin{figure}
\centering
\includegraphics[angle=0.,width=4.0in]{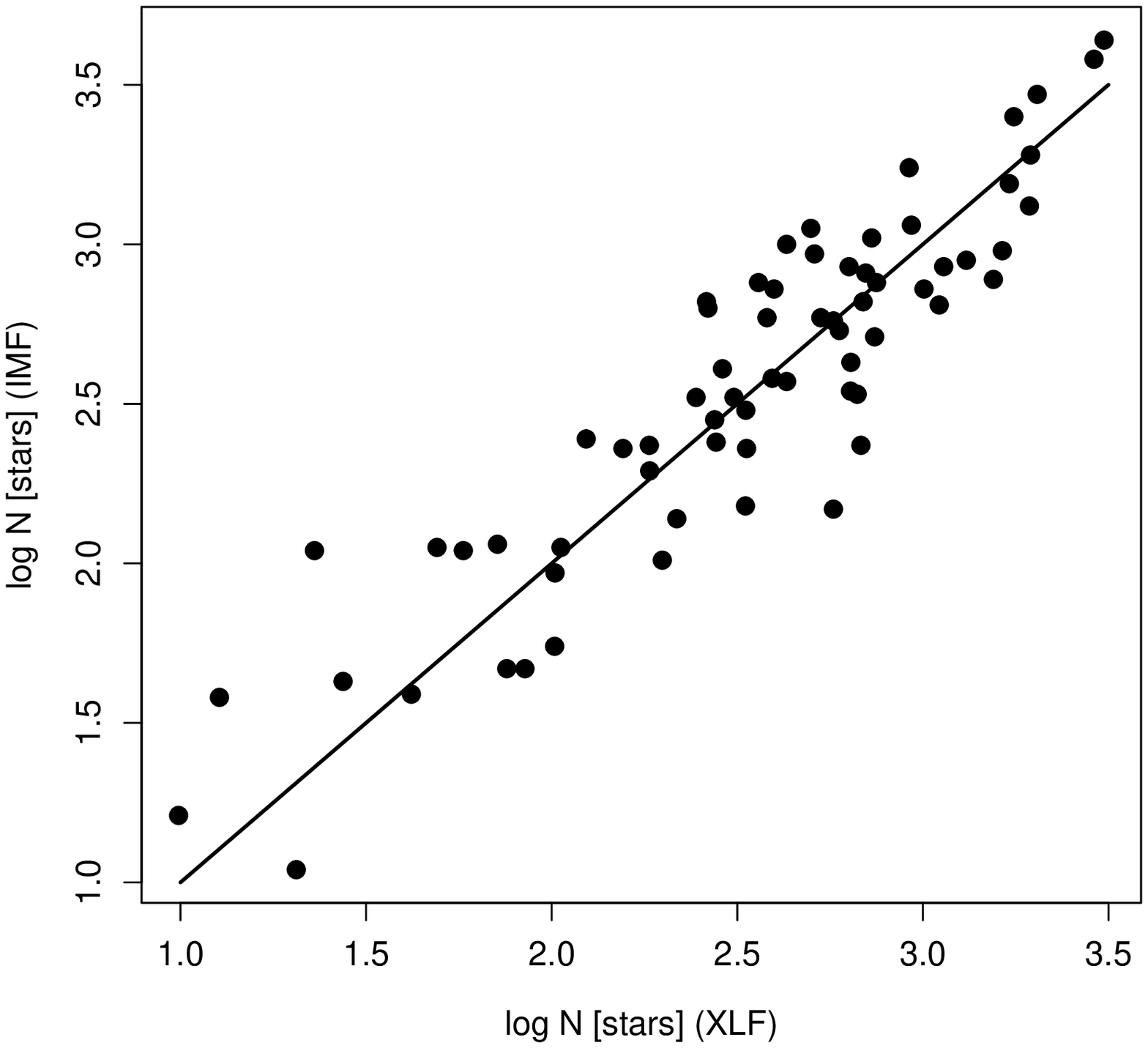}
\caption{Intrinsic numbers of stars within the MYStIX subclusters estimated via the IMF (ordinate) vs.\ XLF (abscissa). The $y=x$ line where both methods produce the same estimate is indicated.
\label{subclust.fig}}
\end{figure}

Figure~\ref{subclust.fig} plots the intrinsic population from IMF analysis vs.\ the intrinsic population from XLF analysis on a log-log plot---the points fall along the $y=x$ line with a root-mean-square of $\sim$0.25~dex, although some points may deviate by up to a factor of $\sim$3. IMF-inferred populations for sparser subclusters are slightly systematically higher than the XLF-inferred populations, but, overall, there is little systematic shift or tilt in the relation. Subclusters that are more highly absorbed tend to have fewer members, but they do not show any more or less deviation from the $y=x$ line than the unabsorbed subclusters. 

\citet{Povich11} find that IR-derived populations estimates, which include the IR excess selected stars that lack detected X-ray counterparts, produce results $\sim$20\% higher than the estimates that just use the X-ray selected sources. However, this offset is small relative to other sources of uncertainty in population estimation, and is not apparent on the log-log plot in Figure~\ref{subclust.fig}. Another source of uncertainty in Figure~\ref{subclust.fig} is the comparison of the completeness limit in $L_{tc}$ to a mass completeness limit. The sample of 839 COUP stars is complete down to 0.1--0.2~$M_\odot$, so there is some uncertainty in determining which mass limit to use for best comparison between XLF and IMF inferred populations. 

\begin{deluxetable}{lrrrrrrr}
\tablecaption{Intrinsic Population Estimates from the XLF and IMF \label{imf_xlf.tab}}
\tablewidth{0pt}
\tablehead{
\colhead{} & \multicolumn{5}{c}{Properties from Paper~I} & \multicolumn{2}{c}{}\\
\cline{2-6}
\colhead{Subcluster} & \colhead{$\alpha$} &\colhead{$\delta$} &\colhead{$r_{4,\mathrm{major}}$} &\colhead{$r_{4,\mathrm{minor}}$} &\colhead{PA} &\colhead{$\log N_\mathrm{XLF}$} & \colhead{$\log N_\mathrm{IMF}$}\\
\colhead{} & \colhead{(J2000)} & \colhead{(J2000)} & \colhead{(arcmin)} & \colhead{(arcmin)} & \colhead{(deg)} & \colhead{(stars)} & \colhead{(stars)}\\
\colhead{(1)} & \colhead{(2)} & \colhead{(3)}& \colhead{(4)}& \colhead{(5)}& \colhead{(6)}& \colhead{(7)}& \colhead{(8)}
}
\startdata
Orion~A  &  83.8110030  &-5.3752777  &0.42  &0.32&  85 & 1.66$\pm$0.13 &\nodata \\
Orion~B  &  83.8154178   & -5.3897248 & 1.93  &1.35  &28  &2.17$\pm$0.10& \nodata \\
Orion~C  &  83.8195378  &-5.3761802 &10.17  &5.19  & 5 & 3.21$\pm$0.02  &3.37$\pm$0.09 \\
Orion~D  &  83.8242661   &-5.2763330  &7.60 & 1.20 & 12 & 1.94$\pm$0.12& \nodata \\
Flame~A  &  85.4270870  &-1.9037960  &5.14  &3.24 &146 & 2.74$\pm$0.03  &2.91$\pm$0.43 \\
W40~A   &  277.8614542  &-2.0940426  &4.54  &4.37 &107 & 2.48$\pm$0.04  &2.66$\pm$0.09 \\
RCW~36~A &  134.8623491 &-43.7555688 & 3.47  &2.32 &122 & 2.73$\pm$0.04  &2.27$\pm$ 0.06 \\
RCW~36~B &  134.8634966 &-43.7571938 & 1.01 & 0.15 & 23 & 1.66$\pm$0.11  &1.94$\pm$0.31 \\
NGC~2264~A & 100.1312445  & 9.8311532 & 1.35 & 1.16 &136 & 1.21$\pm$0.15  &0.94$\pm$0.12 \\
NGC~2264~B &  100.1545919  & 9.7918914 & 0.56 & 0.31& 124 & 1.12$\pm$0.21 &\nodata\\ 
\enddata
\tablecomments{Column~1: Subcluster name from Paper~I. 
Columns~2--3: Celestial coordinates (J2000) for the subcluster center.
Columns~4--5: Semi-major and semi-minor axes for an ellipse 4 times the size of the subcluster core defined in Paper~I. 
Column~6: Position angle of the subcluster ellipse in degrees east from north. 
Column~7: Intrinsic number of stars projected within four subcluster core radii estimated from the XLF analysis. Column~8: Intrinsic number of stars projected within four subcluster core radii estimated from the IMF analysis. A value of ``\nodata'' indicates that a subcluster has too few stars with good $JH$ photometry to estimate a mass-completeness limit and/or the IMF scaling. Estimates of uncertainty include the $\sqrt{N}$ Poisson uncertainty, but do not include the multiple sources of systematic error in the XLF and IMF analysis. 
(A full version of this table is available in the electronic edition of this article---a stub is provided here to show form and content.) 
}
\end{deluxetable}

\clearpage

\section {Intrinsic Stellar Surface Density \label{density.sec}}

\begin{figure}
\centering
\includegraphics[angle=0.,width=5.0in]{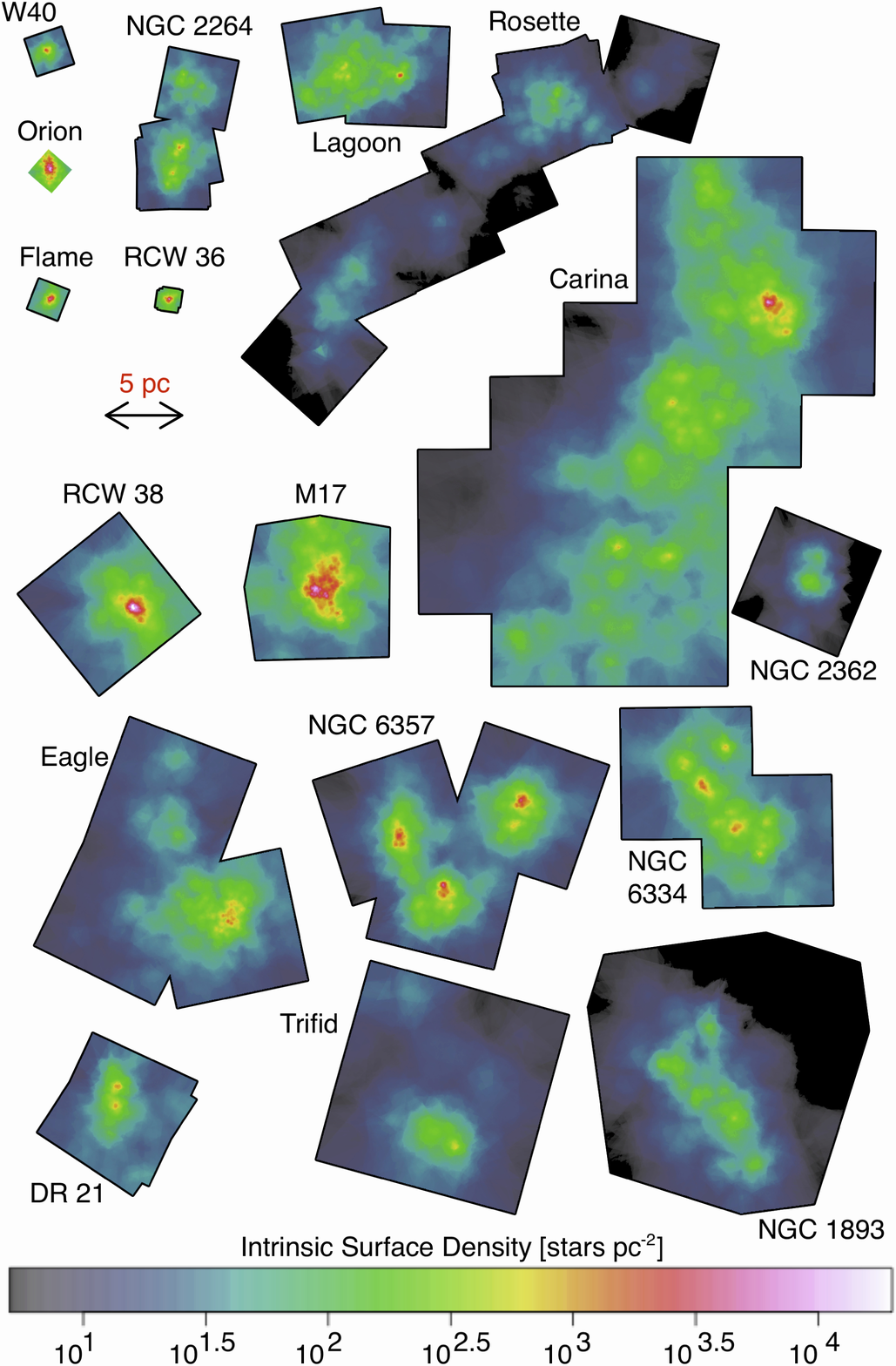}
\caption{Surface densities in all 17 regions, shown using the identical spatial scale (parsec scale given by arrow) and intrinsic surface density scale (stars pc$^{-2}$ scale given by color bar) for each region. (The data behind this figure is provided with the electronic edition of this article.)
\label{all_map.fig}}
\end{figure}

Figure~\ref{all_map.fig} shows intrinsic stellar surface density for the 17 MSFRs regions. Observed surface densities for X-ray selected MPCMs are calculated following the \citet{Ogata03,Ogata04} adaptive-smoothing method and then corrected to the intrinsic populations by dividing the observed surface densities by the detection fraction maps. These maps can thus be directly compared with each other; Figure~\ref{all_map.fig} shows all regions with the same physical length scale \citep[in parsecs based on the distances given by][]{Feigelson13} and the same surface-density units (in stars per square parsec). A 5-pc length scale is drawn. 

The adaptive smoothing method is based on the Voronoi tessellation as implemented by the {\it adaptive.density} function in the {\it spatstat} CRAN package of the R statistical software environment \citep{Baddeley05}. A randomly selected subset of stars are used to create a Voronoi tessellation of the field, which will naturally tend to have smaller cells in regions of higher stellar density. The other stars are used to estimate surface density in these cells. This procedure can be repeated a large number of times using different subsets to create the tessellation and estimate the surface density. To produce our maps, we used a sample containing $N_\mathrm{obs}/5$ stars to create the tessellation and repeated the procedure 100 times, averaging together the results. This method produces results that are similar to other adaptive smoothing methods used by astronomers, such as $k$-th nearest neighbor surface-density estimator \citep{Gutermuth08} or adaptive kernel density estimator (KDE) methods \citep[e.g.,][]{Abramson82}. The choice of what fraction of stars to use to build the tessellation for the Ogata method is analogous to the choice of $k$ for the $k$-th nearest-neighbor method or the kernel size for KDE.

\begin{figure}[h]
\centering
\includegraphics[angle=0.,width=4.0in]{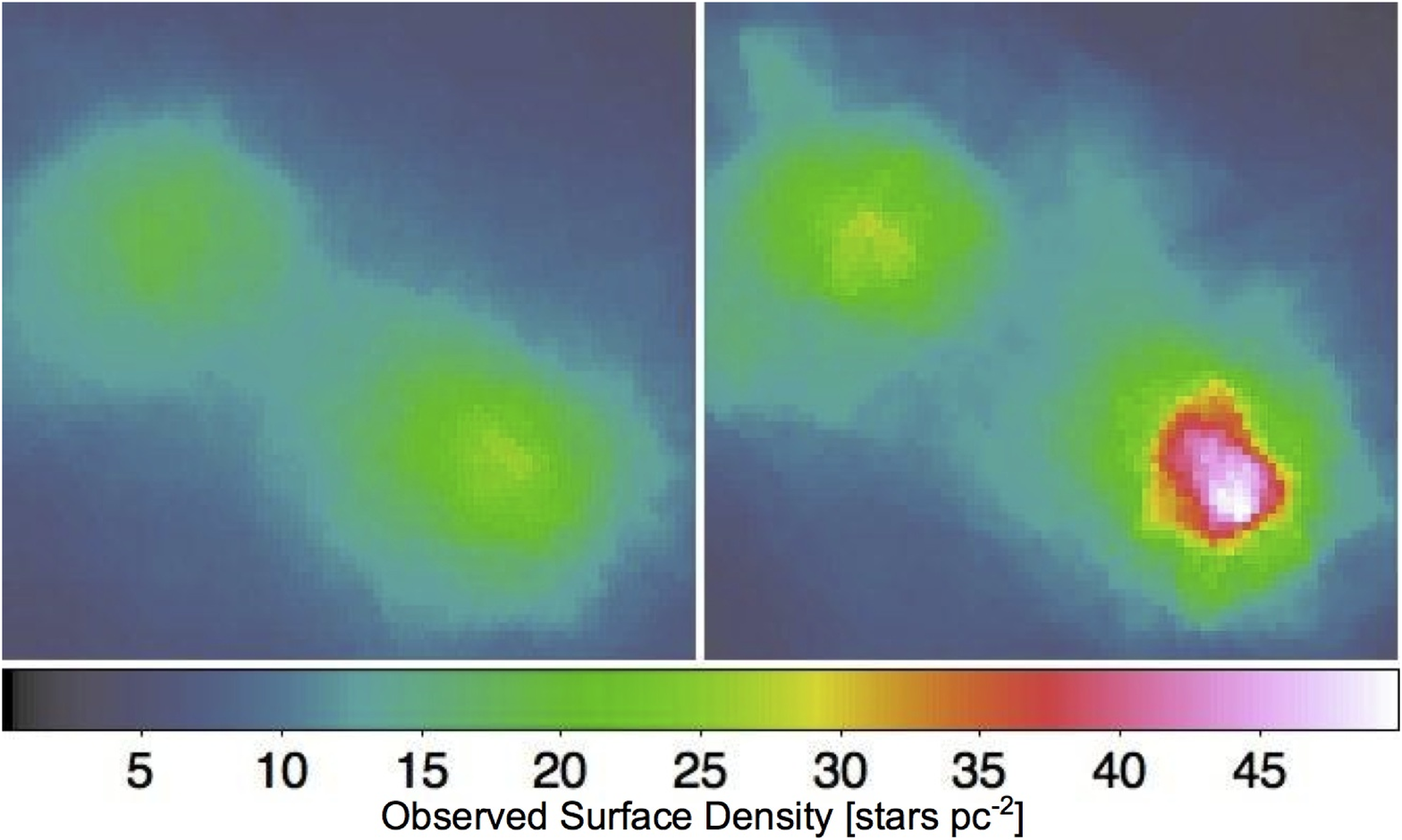}
\caption{Surface density maps for the same region (central part of Trifid) with two different smoothings: in the left panel 10 points are used per Voronoi cell, while in the right panel 5 points are used per Voronoi cell. The color bar shows units of observed stars~pc$^{-2}$.
\label{trifid_smoothing.fig}}
\end{figure}

For any non-parametric smoothing algorithm, there is a tradeoff between bias and variance, with bigger ``kernels'' leading to smaller variance but larger bias. This effect can lead to suppression of peaks in stellar surface density, which is demonstrated by Figure \ref{trifid_smoothing.fig}. When one tenth of the points are used to create the Voronoi tessellation leading to a smoother map (left) and one fifth of the points are used to create the Voronoi tessellation leading to a rougher map (right), the peak in the map is suppressed by a factor of $\sim$2 in the former case compared to the latter case.\footnote{The central subcluster surface densities from Paper~I were obtained by parametric modeling of the unbinned data, so they should not be affected by this bias.} In contrast, surface densities away from local extrema are nearly identical in the two panels. This indicates that these maps (or any other maps of stellar surface density using the various methods listed above) are likely to produce biased values for maximum surface density, but may be reasonably accurate in regions with smooth surface-density gradients.

We have run simulations to test the reliability of the Voronoi surface-density technique, in particular whether stochastic clumping of spatially random points can produce false peaks in the surface-density maps.\footnote{This issue equally affects the $k$-th nearest neighbor and adaptive kernel methods as well.} We generate a point process with complete spatial randomness of 100 points in a unit square and estimate surface density using the Voronoi method with $N_\mathrm{obs}/5$ points to construct the tessellation. This procedure is repeated 10,000 times. We find that the root-mean-square uncertainty in surface density at the location of each point is 0.1~dex, 1\% of points may have surface density values 2 times higher, and 0.01\% of stars may have surface density values 3 times higher. Given that the surface densities in star-forming regions vary over more than 4 orders of magnitude, these stochastic variations are insignificant, and the observed peaks are likely to be real.

\subsection{Descriptions of Surface-Density Maps}

The structures seen in the maps in Figure~\ref{all_map.fig} are the projected distributions of stars in star-forming regions, so physically discrete groups of stars may overlap each other on the map. This is known to be the case for the Orion region---this field of view includes, in order of distance along our line of sight, the periphery of the older NGC~1980 cluster \citep{Alves12}, the ONC, a dense subcluster containing massive stars \citep[BN/KL;][]{Becklin67,Kleinmann67}, and stars embedded in the OMC, including the OMC1-S subcluster \citep{Grosso05}. The clumpiness seen in the \mystix\ maps hints at such structure in other regions. 

The stellar surface density maps in Figure~\ref{all_map.fig} show that most regions have small clumps of stars with extremely high surface density, but most of the area of the star-forming regions have surface densities well below the maxima. The highest peaks in surface density include the core of  the RCW~38 cluster ($\sim$34,000~stars pc$^{-2}$), Orion ($\sim$17,000~stars pc$^{-2}$ for the ONC and $\sim$22,000~stars pc$^{-2}$ for BN/KL), M~17 ($\sim$12,000~stars pc$^{-2}$), the Tr~14 cluster in Carina ($\sim$10,000~stars pc$^{-2}$), and RCW~36 ($\sim$10,000~stars pc$^{-2}$). In the M~17 region, a projected area of 9.8~arcmin$^{2}$ ($= 3.3$~pc$^2$) has surface densities greater than 1000~stars~pc$^{-2}$---substantially larger than for any other \mystix\ MSFR. While, at the other extreme, the Rosette Nebula has an overall low stellar surface density.

The ONC plays a paradigmatic role in our understanding of young stellar clusters \citep[e.g.,][]{Hillenbrand97,Getman05,Bally00}, and the surface-density maps show that it is similar in size and density to many of the densest clumps of stars in other MSFRs, even regions in which a much larger physical area has been surveyed. An example of this is the NGC~6357 region, which has three dense clusters similar to the ONC. Nevertheless, not all of the \mystix\ MSFRs regions contain structures comparable to the ONC; neither the Eagle Nebula nor the Lagoon Nebula are as centrally concentrated as the ONC, despite having larger total young stellar populations. Older MSFRs like NGC~1893 and NGC~2362 entirely lack dense cluster cores.

The simulation from Section~\ref{strata.sec} of the degradation on the Orion Nebula data if this region were observed with the distance and exposure time for Carina (reducing the number X-ray sources from 1216 to 120) reveals how results for more distant \mystix\ regions might be affected by lower sensitivity. The smoothed surface-density map for the simulated observation has a broadened central core, and, hence, the maximum surface density at the center of the ONC is reduced. Most of the highly embedded stars around BN/KL were removed, so this subcluster is not seen. However, the range of surface densities outside these peaks is not changed significantly. Thus, in other MSFRs, one may expect that some of the small but dense subclusters may be missing from our surface-density maps.

\section{Histograms of Surface Density}

The surface density at the location of a star, $\Sigma_\star$, can provide useful information about the environment that young stars experience in star-forming regions \citep[e.g.,][]{Lada03,Gutermuth05,Jorgensen08}. Dynamical equilibration and violent few-body interactions, for example, will occur only in dense stellar cores.  Low density regions, in contrast, can produce dynamically fragile wide binary systems \citep[e.g.,][]{Feigelson06}. For each of the $\sim$17,000 MPCM stars in our sample, $\Sigma_\star$ is interpolated from the intrinsic surface-density maps. For each MPCM in this sample, there are an average of ($1/f_\mathrm{full} - 1$) undetected young stars at nearly the same location. Therefore, to estimate the intrinsic distribution of $\Sigma_\star$ for a star-forming region using a histogram, the number of stars in each bin is calculated by weighting a star at location $(\alpha,\delta)$ by $1/f_\mathrm{full}(\alpha,\delta)$. Thus, the summation of values in all bins should equal the inferred total number of stars in a region, $N_\mathrm{tot}$, rather than the number of observed stars, $N_\mathrm{obs}$.

\begin{figure}
\centering
\includegraphics[angle=0.,width=6.5in]{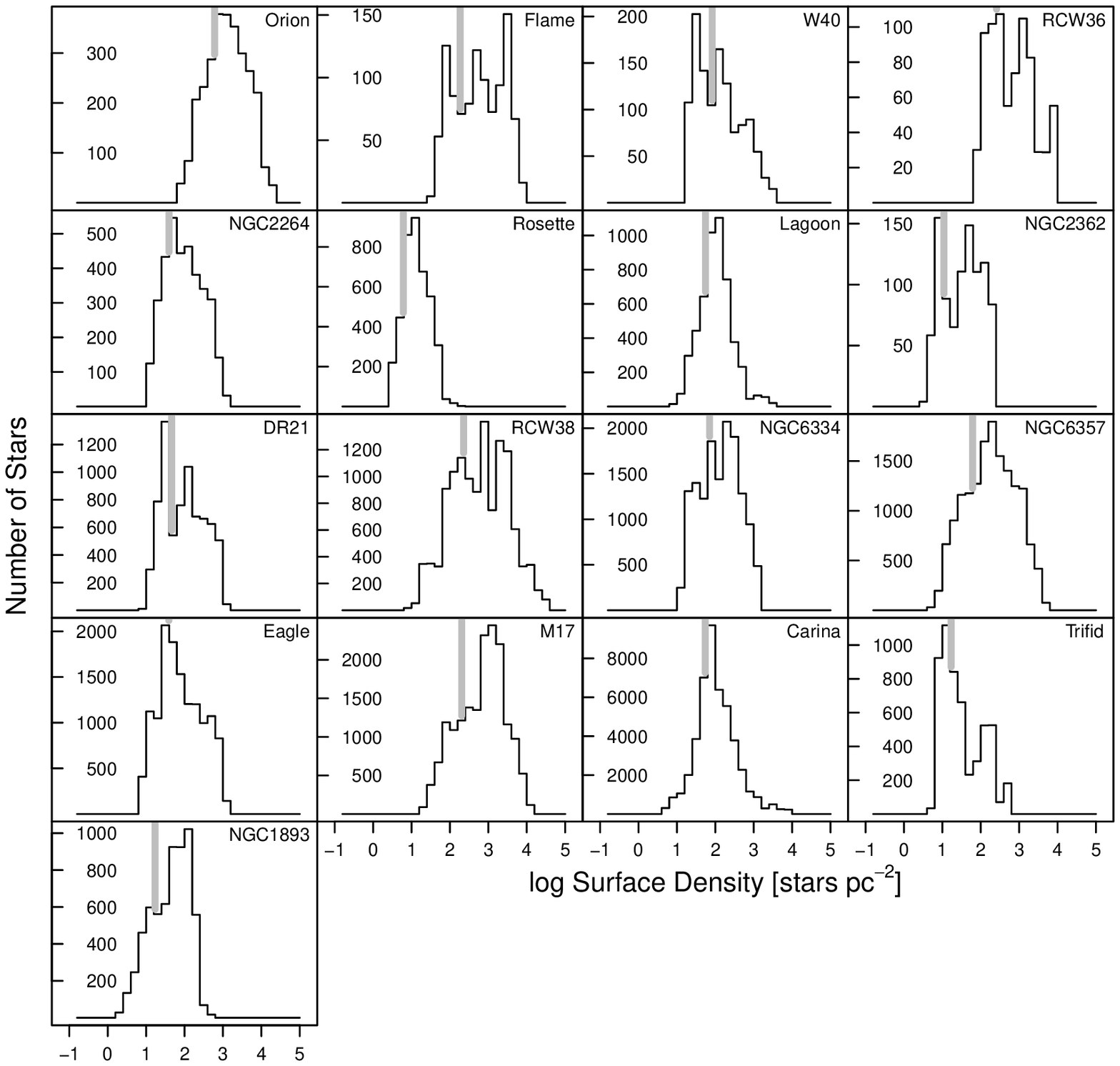}
\caption{The surface density distributions for each individual star-forming region. The histogram bin widths are 0.2~dex, and the $x$ axes are the same for each plot. The gray lines indicate the average surface density in the field of view, defined as the total number of stars divided by the area of the field of view. Left to right and top to bottom: Orion, Flame, W~40, RCW~36, NGC~2264, Rosette, Lagoon, NGC~2362, DR~21, RCW~38, NGC~6334, NGC~6357, Eagle, M~17, Carina, Trifid, and NGC~1893.
\label{hist.fig}}
\end{figure}

Figure~\ref{hist.fig} shows these $\Sigma_\star$ histograms for each of the 17 MSFR, with a bin size of 0.2~dex.  
These graphs show that $\Sigma_\star$ ranges over $\sim$4 orders of magnitude from $\Sigma_\star\sim1$ to 30,000~stars~pc$^{-2}$. The majority of the young stars in the \mystix\ survey lie in regions with $\Sigma_\star = 10$--10,000~stars~pc$^{-2}$.

Comparison of the local surface densities in the neighborhoods around stars to the average surface density across the entire field of view is useful for quantifying the degree of local clustering.
This is the principal behind the nearest-neighbor test for clustering in spatial-point patterns by \citet{Diggle83}, \citet{Ripley88}, and \citet{Cressie91}. The thick, gray lines superimposed on each histogram in Figure~\ref{hist.fig} show the average surface density over the entire field of view. For every \mystix\ region, the median surface density at the location of stars is greater than the average surface densities across the region by 40--400\%, indicating strong clustering in all cases. In addition, all regions have at least a few stars in subregions with surface densities 1--2 orders of magnitude above the average for the field of view.

There is much variety in the positions of the histogram peaks (i.e.\ the mode of the distribution). Orion peaks at the highest $\Sigma_\star$ for the \mystix\ sample (1000~stars~pc$^{-2}$), while the peak for Trifid is the lowest (10~stars~pc$^{-2}$). The distributions of $\log \Sigma_\star$ are often asymmetric around the peak. Some regions have a narrower distribution of $\log \Sigma_\star$, like Rosette, Lagoon, and Carina, while others have a wider distribution, like Flame, RCW~38, NGC~6357, and Eagle. A few regions have statistically significant multimodality, including NGC~2362 and Trifid, where the null-hypothesis of a unimodal distribution is rejected by Hartigan's dip test\footnote{
The hypothesis test for multimodality from \citet{Hartigan85} is implemented in the {\it diptest} CRAN package of the R statistical software environment \citep{Maechler13}.
} 
at $p<0.01$. For Trifid, the different modes correspond to different subclusters from Paper~I, which have significantly different mean densities allowing them to be distinguished on this diagram \citep[cf.][]{Pfalzner12}. The multimodal structure that appears to exist in the histograms of Flame, RCW~36, and RCW~38 is only marginally significant ($p\approx0.05$).

There are limitations to this survey of $\Sigma_\star$ due to the small FOVs of the \mystix\ project, given that the nearest Galactic star-forming complexes can cover many square degrees on the sky. It is important to note that the average surface densities for different regions cannot be directly compared with each other because they depend on the size of the field of view; for example, the small field of view for Orion only captures the dense region around the ONC, while the large field of view for Carina captures a much wider variety of environments. This field-of-view selection effect will also influence values of $\Sigma_*$, and must be carefully taken into consideration when comparing surface densities in two different regions. The less active sites of star formation in large star-forming complexes may often lie outside the \mystix\ fields of view, so the \mystix\ survey does not represent star-formation in low surface-density regions particularly well.  The simulated reduced-sensitivity Orion data (Section~\ref{strata.sec}) also allows us to examine the effect of lower source-detection rates for more distant regions with shorter \Chandra\ observations. The histogram produced from the 120 sources in this sample (not shown) spans a $\Sigma_\star$ range from $100$--$6000$~stars~pc$^{-2}$, with a peak just under 1000~stars~pc$^{-2}$. Although, the maximum $\Sigma_\star$ is reduced, the mode of the distribution is nearly identical to the original data.

\clearpage\clearpage

\subsection{Comparison of Surface Density Distributions in \mystix\ and B10}

Care must be taken when comparing the results here with the results from B10 due to different selection effects on the young stellar samples used for each study. The sky coverage for both \mystix\ and B10 is the result of multiple projects with different objectives, so neither represents an unbiased survey of star-forming environments in the Galaxy. Perhaps most important, the B10 survey is based only on infrared-excess stars, and thus misses the large X-ray selected disk-free population.  In MPCM samples, the stars with no detected IR excess typically outnumber the disk-bearing subpopulation by 2-3:1, and sometimes more than 7:1 \citep[][their Table~1]{Broos13}.  It is therefore not surprising that stellar surface densities derived from the MYStIX survey are higher than those derived by B10.  

One of the main observations of B10 is that the $\Sigma_\star$ distribution of their sample has no discrete modes (peaks in the logarithmically binned histogram) corresponding to ``distributed'' star formation or ``clustered'' star formation. As a result, they concluded that the histogram of surface densities alone is not sufficient for determining which stars are clustered and which stars are not. This assertion is supported by \citet{Gieles12} who demonstrate that even if all stars are born in clusters, cluster expansion can yield a variety of different surface density distributions, including the log-normal distribution of B10. \citet{Pfalzner12} additionally demonstrate that for \citet{King62} cluster density profiles, the low density portion of the cluster outside the dense core can make it difficult to identify distinct peaks in the $\Sigma_\star$ distribution, even if they do exist. Our results  show that different MSFRs have radically different $\Sigma_\star$ distributions (Figure~\ref{hist.fig}), rather than following a universal distribution. It is thus difficult or impossible to identify any dividing surface density threshold from the empirical $\Sigma_\star$ data. B10 also see shifts in the peaks of the distributions for the different surveys they use, $\sim$7~stars~pc$^{-2}$ for the Gould Belt+Taurus and $\sim$30~stars~pc$^{-2}$ for the c2d regions. 

Another conclusion of B10 is that the unimodal log-normal distribution---peaked at 22~stars~pc$^{-2}$ with standard deviation width of 0.85~dex---means that most stars form outside clusters, while densities like 100, 1000, 10,000~stars~pc$^{-2}$ are in the tail of the distribution.\footnote{Nevertheless, B10 note that they are not sensitive to the extreme high $\Sigma_\star$ ``tail'' for the ONC.} They choose a surface-density threshold of 200~stars~pc$^{-2}$ to be their definition of a cluster because this is where stars become likely to interact with their neighbors. \citet{Gieles12} and \citet{Pfalzner12} caution about using projected surface-densities to estimate the fraction of stars undergoing local dynamical interactions because analysis of the empirical surface-density histograms for a sample does not take into account cluster evolution, radial cluster structure, and superposition of discrete clusters. Thus, they argue that stellar interaction could be important even if the empirical histograms show most stars lie in environments below an astrophysically selected threshold like 200~stars~pc$^{-2}$.

\begin{figure}
\centering
\includegraphics[angle=0.,width=6.5in]{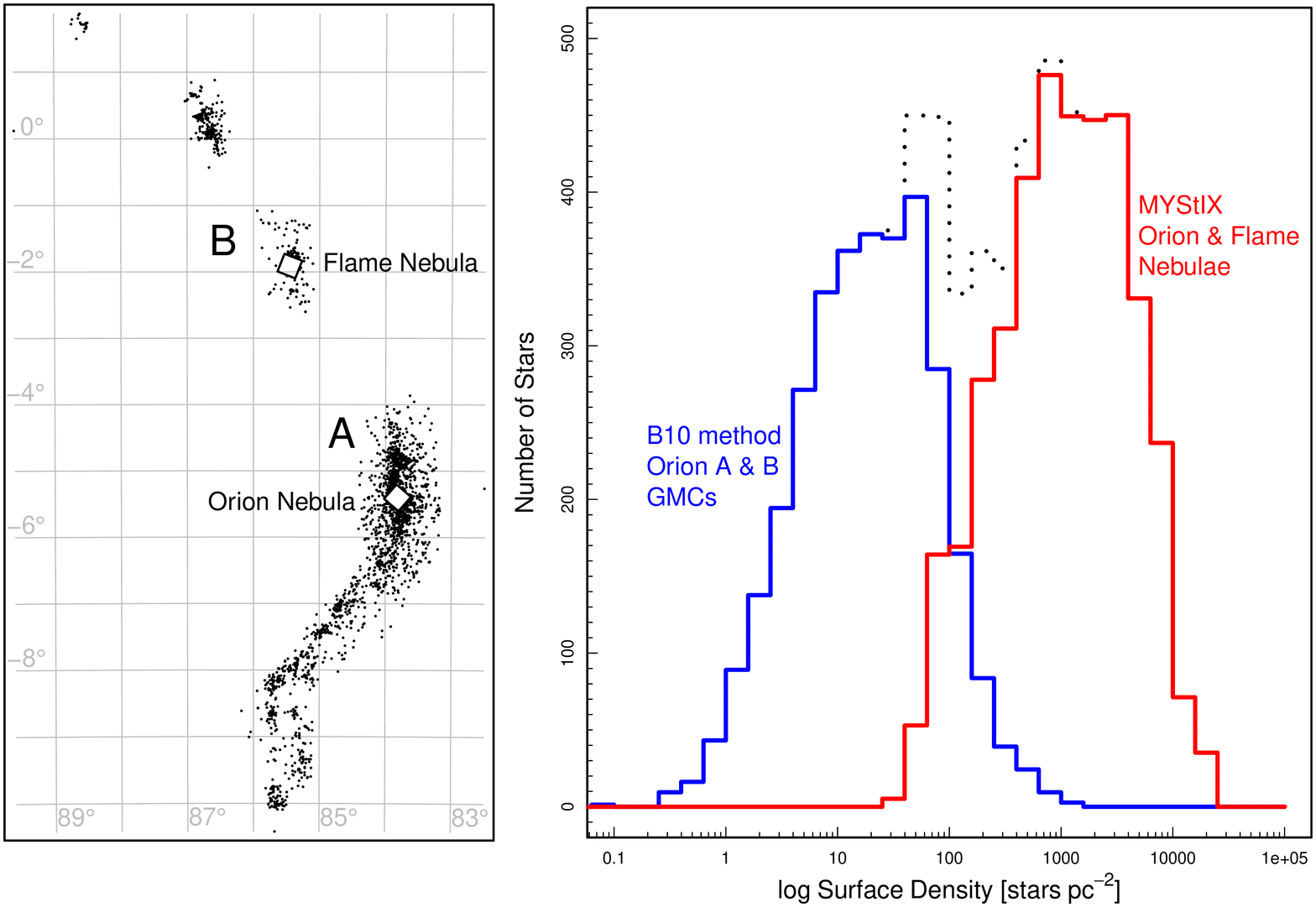}
\caption{Left: Points mark the positions of YSOs from \citet{Megeath12} in the Orion A and B molecular clouds, which meet the B10 selection criteria and lie outside the Orion Nebula and Flame Nebula \mystix\ fields (indicated by polygons). Right: The $\Sigma_\star$ histograms for B10 objects and \mystix\ objects are shown by blue and red lines, respectively. The coadded B10+\mystix\ histogram is shown with a dashed line. 
\label{orion_compare.fig}}
\end{figure}

Both \mystix\ and B10 investigate star-formation in the Orion Giant Molecular Clouds, including the Orion~A and~B molecular clouds, containing the Orion Nebula and Flame Nebula, respectively, making this a useful region to directly compare results. 
The combination of the B10 and the \mystix\ surface-density distributions for this star-forming region provides a less biased estimate of the full surface-density distribution. 
We have emulated the B10 analysis for the larger-scale Orion star-forming region in the areas outside the \mystix\ fields of views, using the \citet{Megeath12} catalog of YSOs, with a magnitude limit at IRAC $[3.6]=14$~mag. The objects excluded due to overlap with the \mystix\ fields of view include 19\% of the B10 Orion sample.
The distribution of these sources on the sky is shown in the left panel of Figure~\ref{orion_compare.fig}, with polygons cut out representing the Flame Nebula \mystix\ field of view in the north and the Orion Nebula \mystix\ field of view in the south. 
To compute the $\Sigma_\star$ histogram for the Orion A and B region (excluding the MYStIX fields for the Orion and Flame nebulae), we follow B10 by assuming a uniform disk fraction of 65\%, so all densities and bin heights are increased by a factor of 1.35. 

The right panel of Figure~\ref{orion_compare.fig} shows these two histograms, labeled ``B10 method'' for the Orion A and B molecular clouds (excluding the \mystix\ Orion and Flame Nebulae fields of view) and labeled ``\mystix'' for the coadded Orion Nebula and Flame Nebula histograms. Both methods attempt to account for intrinsic numbers of stars: in B10's case, by accounting for missing Class III stars with a uniform correction; in our case, through the XLF and IMF methods described in Section 3. At 5~Myr and a distance of 0.414~kpc, a PMS star of 0.1~$M_\odot$ would have an $L$-band ($\approx$~the IRAC [3.6] band) photospheric magnitude of $\sim$13~mag \citep{Siess00}; so B10's sample should be sensitive to YSOs down to the hydrogen-burning mass limit. Although the ``B10 method'' histogram is missing 19\% of the  B10 stars due to overlap with the \mystix\ regions, the location of its peak is consistent with the histogram in Figure~1 of B10.

The full distribution for the Orion Giant Molecular Clouds in Figure~\ref{orion_compare.fig} is clearly very different from the results in Figure~1 of B10. Furthermore, based on the total numbers of stars in these histograms, clustered stars make up one of the dominant components of the $\Sigma_\star$ distribution, rather than just being an extreme value ``tail'' of the distribution. Underestimation of the number of stars in high density regions was also commented on by \citet{Pfalzner12}.
The coadded B10 and \mystix\ histograms (the dashed line) has some bimodal structure, with peaks around 100 and 1000~stars~pc$^{-2}$ and a shoulder around 20~stars~pc$^{-2}$, but this may be an artefact of selection effects in both studies. 

\begin{figure}
\centering
\includegraphics[angle=0.,width=6.5in]{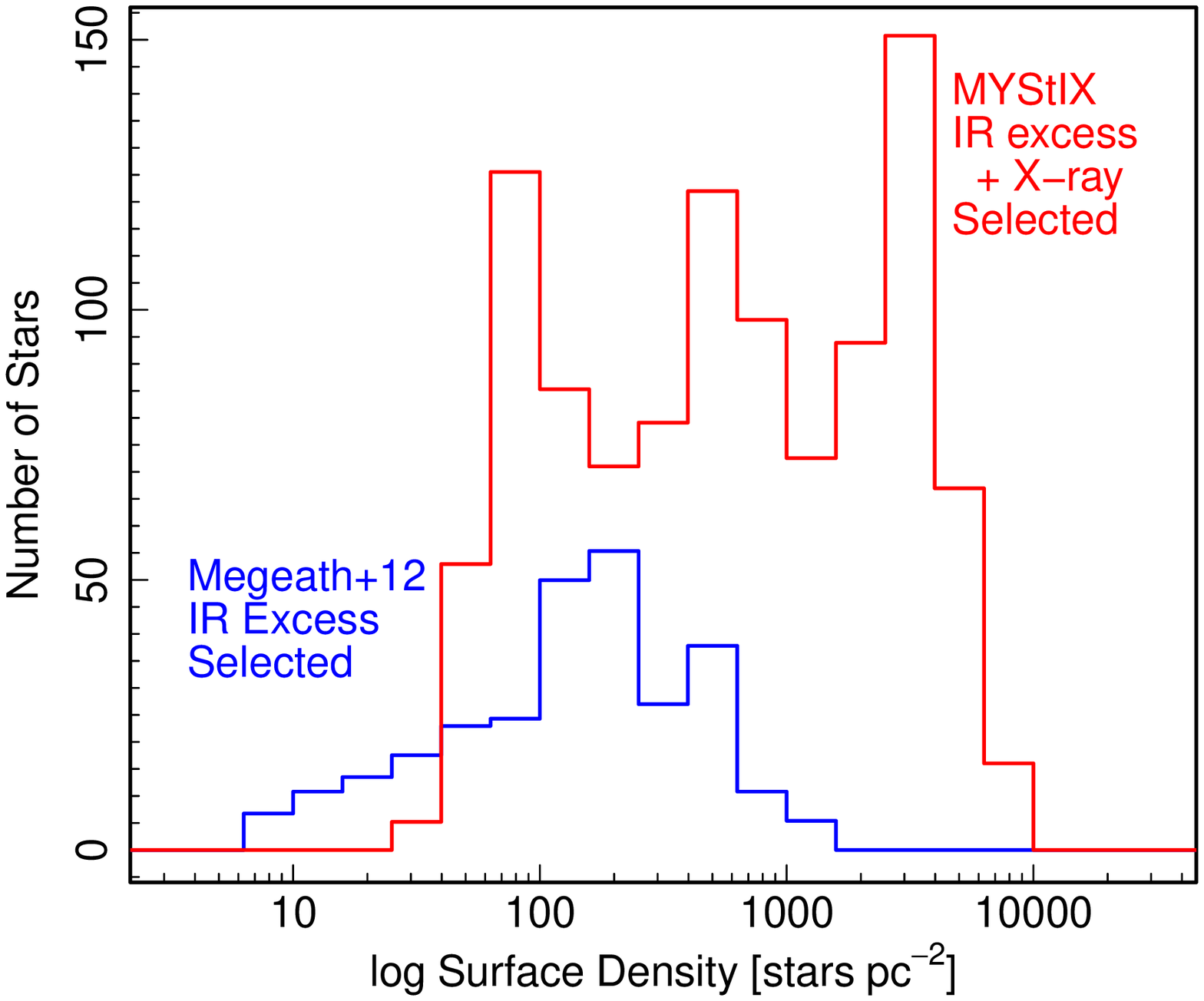}
\caption{
 For the $17^\prime\times17^\prime$ \mystix\ field of view for the Flame Nebula, $\Sigma_\star$ distributions are shown which are inferred using only IR-excess selected members (B10 method; blue histogram) and IR-excess plus X-ray selected members (\mystix; red histogram). 
\label{flame_compare.fig}}
\end{figure}

Figure~\ref{flame_compare.fig} shows $\Sigma_\star$ histograms for just the Flame Nebula region, which are obtained using the B10 method with only IR-excess selected members from \citet{Megeath12} and the histogram obtained from \mystix\ using both IR-excess and X-ray selected members. Visual inspection of the \Spitzer\ IRAC images reveals that IR sources are likely lost in the wings of bright sources and in regions with bright nebulosity. As a result, the IR excess only method underestimates both the number of stars in this cluster and the stellar surface densities.

Limitations that affect the IR-excess census of YSOs (outside the dense Flame Nebula and Orion Nebulae) include variations in \Spitzer\ sensitivity due to the presence of nebulosity \citep[e.g.,][]{Muench07} and variations in the disk fraction \citep[e.g.,][]{Getman09,Kuhn10,Povich11,Getman12,Getman14a}. If a significant numbers of young stars were missed in the outer portions of the star-forming complex, it might increase the low-density component of the combined histogram to the point where it is comparable to the high-density component. Results from a large XMM survey of the Orion~A molecular cloud \citep[e.g.,][]{Pillitteri13} will be able to address some of these limitations. Limitations to the \mystix\ censuses include reduced sensitivity of X-ray selection to protostars \citep{Prisinzano08} and highly absorbed clumps of stars and assumptions about a uniform XLF and IMF.

While improved information about the typical environments in which stars form would be useful for theoretical models of star-formation and cluster formation and evolution; it is difficult to generalize results like the ones here or from B10 to star-formation in the Galaxy as a whole. Even within the \mystix\ sample, surface density distributions from one region provide little information about other regions (Figure~\ref{hist.fig}). We have demonstrated that B10's results are likely not valid for the Orion molecular clouds ($\sim$70\% of their sources). But their results are likely to be reasonable for other nearby star-forming clouds that are less affected by crowding and nebulosity, although scalings to include disk-free subpopulations may vary. Nevertheless, the high surface densities seen in other MSFRs indicate that the results for the Orion molecular clouds are not anomalous, and may indeed be typical for most stars, in the context of Galactic star-formation. The initial cluster mass function described by a power-law with index of approximately -2 over a large cluster mass range \citep{Lada03,Chandar10,PortegiesZwart10,Lada10,Ryon14,Fouesneau14}, would imply that stars are roughly equally likely to form in complexes of different masses. This would mean that star formation is evenly spread between small Taurus/Chamaelion-type clouds with tens-to-hundreds of stars, smaller OB star-forming regions like NGC~2264 containing hundreds to thousands of stars, giant molecular clouds like the Orion compelex containing thousands to $10^4$ stars, and major star-forming complexes like Carina containing $10^4$--$10^5$ stars. Since massive stars appear in the second group, most stars are likely born in regions containing or influenced by massive stars.

\clearpage\clearpage

\clearpage\clearpage

\section{Models of Subclustered and Unclustered Stars}

In Paper~I, we used ``finite mixture models'' to subdivide the observed populations of stars in the \mystix\ MSFRs into statistical clusters and ``unclustered'' uniformly distributed components.\footnote{There are multiple different definitions for ``young stellar cluster'' in the literature, and here we use cluster in the statistical sense. } 
This is a common method of cluster analysis which provides ``soft'' cluster assignments for individual stars---probabilities that a star belongs to a particular subpopulation. Our models, which estimate the spatial surface density in star-forming regions, are the composite of multiple parametric models for each of the clustered and unclustered subpopulations. The subclusters were modeled using ellipsoidal surface density profiles, similar to the model forms investigated by \citet{Hillenbrand98} and \citet{Pfalzner12}, and an additional spatially-constant component was added to account for stars that are not part of any subcluster. The number of subclusters, the subcluster parameters, and the number of stars in each component were determined by model fitting with model selection based on the widely-used Akaike Information Criterion penalized likelihood measure. This method requires a parametric form to be assumed for the subcluster models, and assumes that stars which are not part of subclusters are uniformly distributed, which are not necessarily accurate assumptions; however, the best-fit results in Paper~I show that these models are able to reproduce the observed surface density distributions with low-amplitude residual maps. 

Table~\ref{uncl.tab} gives the intrinsic populations for the finite-mixture-model ``unclustered'' component. Corrections for sample incompleteness of this component were obtained using the same methods described in Section~\ref{detection.sec} to obtain intrinsic populations for subclusters in Table~\ref{imf_xlf.tab}. The number of unclustered stars from our analysis differs from the results of \citet{Feigelson11} in Carina and \citet{Feigelson09} in NGC~6334 because those studies use a threshold method, while we use the mixture model method. Table~\ref{uncl.tab} also reports the fraction of stars belonging to the unclustered component.

In the models 80--90\% of the young stellar populations in the \mystix\ fields of view are members of subclusters, while 10--20\% of the young stellar populations are part of a component that is approximately uniformly distributed. The stars that are part of the latter component could be made up of stars that formed in relatively isolated parts of a MSFR, an earlier generation of star-formation that has been dispersed, stars that drifted away from subclusters, or clumps of stars with too few members to be identified as subclusters. 
Some of the complexes with larger fractions of unclustered stars include younger regions with more active star formation, for example W~40 (20\% unclustered), NGC~2264 (18\% unclustered), and the Rosette Molecular Cloud (22\% unclustered). This may be the result of young stars forming in networks of molecular filaments outside the main cluster \citep[e.g.,][]{Andre10}, which could later become incorporated into the cluster in a hierarchical merger scenario \citep{Maschberger10}. In contrast, for DR~21 (13\% unclustered) the youngest objects tend to be embedded in the dense filament passing through the center of the cluster, while the unclustered objects have an older median age \citep{Getman14b}. The consistently low fraction of stars in the unclustered components of the \mystix\ regions highlights that the distribution of stars is very clumpy, rather than spatially smooth.

The fraction of stars in subclusters, combined with information about subcluster properties (whether or not the \mystix\ subclusters are gravitationally bound is investigated in Paper~III), will help characterize the future evolution of the MYStIX MSFRs. The fraction of stars that are born into groups that are initially gravitationally bound has been related to the fraction of star formation that results in bound open clusters \citep{Kruijssen12a}.

\begin{deluxetable}{lrrr}
\tablecaption{Clustered and Unclustered Young Stellar Populations \label{uncl.tab}}
\tablewidth{0pt}
\tablehead{
\colhead{Subcluster} & \colhead{$N_\mathrm{uncl}$} & \colhead{$N_\mathrm{clust}$}& \colhead{\% clustered}\\
\colhead{} & \colhead{(stars)} & \colhead{(stars)} & \colhead{(\%)}\\
\colhead{(1)} & \colhead{(2)} & \colhead{(3)} & \colhead{(4)}
}
\startdata
Orion 	& 170	 & 2400	 &	93 \\
Flame 	& \nodata	 & 	800 &	\nodata \\
W~40 	&	100 & 	420 &	80 \\
RCW 36 	&	35 & 	510 &	93 \\
NGC 2264 &	340 & 	1600 &	82 \\
Rosette 	&	560 & 	1900 &	78 \\
Lagoon 	&	\nodata & 3800	 &	\nodata \\
NGC 2362 &	26 & 	570 &	96 \\
DR 21 	&	370 & 	2500 &	87 \\
RCW 38 	&	\nodata & 	9900 &	\nodata \\
NGC 6334 &	760 & 	8600 &	92 \\
NGC 6357 &	\nodata & 	12000 &	\nodata \\
Eagle 	&	\nodata & 	8100 &	 \nodata \\
M 17 	&	\nodata & 	16000 &	\nodata  \\
Carina 	&	2700 & 	31000 &	91 \\
Trifid 	&	360 & 	2700 &	88 \\
NGC 1893 &	67 & 	4500 &	 98 \\
\enddata
\tablecomments{Properties of total intrinsic subclustered and unclustered young stellar populations. Column~1: Region name. Column~2: Intrinsic number of young stars in the unclustered component (integrated over the whole field of view). Column~3: Sum of intrinsic subcluster populations for a region (integrated over the whole field of view). Column~4: Fraction of total young stellar population in subclusters. 
}
\end{deluxetable}

\clearpage\clearpage

\section{Summary}

We present total intrinsic populations and their surface-density distributions for 17 massive star-forming regions using catalogs of young stars from the \mystix\ project. 

\begin{itemize}

\item The analysis of X-ray luminosities of stars in 17 star-forming regions shows results that are consistent with the hypothesis of a ``universal XLF'' for PMS stars. Figure~\ref{full_xlfs.fig} displays the cumulative distributions of 0.5--8.0~keV, absorption-corrected X-ray luminosities for the $\sim$17,000 MPCM stars in our sample. The ONC XLF from the COUP study is still the only XLF in our study complete down to 0.1--0.2~$M_\odot$. However, the other regions show close agreement with the ONC over the range of X-ray luminosities for which their intrinsic XLF is observed. 

\item The COUP XLF and standard near-IR Initial Mass Functions are used to extrapolate total intrinsic numbers of stars in the MYStIX fields of view, which are given in Table~\ref{full_xlfs.tab} (Column~3). The total intrinsic numbers of stars in individual subclusters from Paper~I are provided in Table~\ref{imf_xlf.tab}. About $\sim$16\% of the full population appears in the \mystix\ samples, but detection fractions vary from region to region.

\item There is consistency between total number of stars calculated from XLF extrapolation and calculated from IMF extrapolation (Figure~\ref{subclust.fig}). There is little systematic offset in total populations calculated by these two methods, but the root-mean-square scatter in the relation is 0.25~dex. This result validates the use of the XLF to infer intrinsic total populations of stars.

\item Intrinsic stellar surface-density maps are provided for 17 star-forming regions, plotted using the same physical scale and surface density scale (Figure~\ref{all_map.fig}). This set of maps provides one of the best and least biased views available today of the spatial distributions of stars formed in giant molecular clouds over the past few million years. Stars are highly clustered within these fields, with subclusters of stars similar in size and density to the ONC existing in several of the complexes. These surface density maps are available in FITS format in the online edition of this paper.

\item The highest surface densities in the \mystix\ regions are found in the core of RCW~38 ($>$30,000~stars~pc$^{-2}$). Another notable high--surface-density star-forming region is M~17, which has an unusually large area with $>$1000~stars~pc$^{-2}$. The other regions containing stars in similar or greater numbers to M~17 (NGC~6357 and Carina) have much smaller high--surface-density regions.

\item \mystix\ finds surface densities in massive star-forming regions ranging from $\sim$1 to $\sim$30,000~stars~pc$^{-2}$, exceeding the surface density of the ONC Trapezium core by a factor of 2. Peaks in the logarithmically binned histograms of surface density vary from region to region, with some peaked near 22~stars~pc$^{-2}$ similar to B10's results (e.g., Rosette), but with others having surface-density modes between 200~stars~pc$^{-2}$ (e.g., NGC~1893) and 2,000~stars~pc$^{-2}$ (e.g., M~17). The variation in the shape of these distributions indicates that there is no universal surface-density distribution applicable to all types of star-forming regions and no special value of surface density characteristic of Galactic star formation. 

\item In the \mystix\ regions, more than half of the young stars lie in regions with high surface densities with 100-10,000~stars~pc$^{-2}$. Given that most stars form in star-forming regions containing O-type stars, this result suggests that a large fraction of stars in the Galaxy formed in such dense environments. Nevertheless, a quantitative determination of how great this contribution would be would require an unbiased survey of many different star-forming environments.
We also demonstrate that for the combined young stellar population of the Orion molecular clouds, including $\sim$70\% of the stars used in B10's study, the peak of the logarithmically binned distribution is $>$1000~stars~pc$^{-2}$, not $\sim$20~stars~pc$^{-2}$ as they find for their subsample (Figure~\ref{orion_compare.fig}). 

\item Within the \mystix\ fields of view, 80--90\% of stars belong to subclusters identified in Paper~I and 10--20\% of stars belong to the uniformly distributed component. These values, combined with information about gravitational boundedness of subclusters, have implications for cluster survival models.

In theoretical studies of star formation, it is important not to underestimate the number of stars born in high-density regions where interactions between stars and cluster dynamics can have important effects on binary distributions, few-body stellar interactions, and cluster survival. We have demonstrated that X-ray selection of young stars can be very helpful in this regard, detecting several times more stars than samples based on IR-excess disks and allowing more accurate estimation of the populations of dense clusters. 


\end{itemize}

\acknowledgements Acknowledgments:  

The \mystix\ project was supported at Penn State by NASA grant NNX09AC74G, NSF grant AST-0908038, and the \Chandra\ ACIS Team contract SV4-74018 (G.~Garmire \& L.~Townsley, Principal Investigators), issued by the \Chandra\ X-ray Center, which is operated by the Smithsonian Astrophysical Observatory for and on behalf of NASA under contract NAS8-03060.  M.A.K. also received support from NSF SI2-SSE grant AST-1047586 (G. J. Babu, PI), a CONICYT Gemini grant (J. Borissova, PI) from the Programa de Astronom\'{i}a del DRI Folio 32130012 (Fondecyt Regular No.\ 1120601), and the Chilean Ministry of Economy, Development, and Tourism's Millennium Science Initiative through grant IC12009,  awarded to Millennium Institute of Astrophysics, and Fondecyt project number 3150319. We thank Susan Pfalzner, Leisa Townsley, Patrick Broos, G. Jogesh Babu, Kevin Luhman, and Richard Wade for providing advice on this paper. The anonymous referee also gave thoughtful and useful comments. This research made use of data products from the \Chandra\ Data Archive. This work is based on observations made with the {\it Spitzer Space Telescope}, obtained from the NASA/IPAC Infrared Science Archive, both of which are operated by the Jet Propulsion Laboratory, California Institute of Technology under a contract with the National Aeronautics and Space Administration. This research has also made use of SAOImage DS9 software developed by Smithsonian Astrophysical Observatory and NASA's Astrophysics Data System Bibliographic Services.

\end{document}